\documentclass[10pt,preprint]{aastex}

\shorttitle{ALFV\'EN WAVE SOLAR MODEL}
\shortauthors{van der Holst et al.}

\begin{document}

\title{ALFV\'EN WAVE SOLAR MODEL (AWSOM): CORONAL HEATING}

\author{B. van der Holst\altaffilmark{1},
  I.V. Sokolov\altaffilmark{1},
  X. Meng\altaffilmark{1},
  M. Jin\altaffilmark{1},
  W.B. Manchester IV\altaffilmark{1},
  G. T\'oth\altaffilmark{1}, T.I. Gombosi\altaffilmark{1}}

\altaffiltext{1}{Atmospheric Oceanic and Space Sciences,
  University of Michigan, Ann Arbor, Michigan 48109}

\begin{abstract}

We present a new version of the Alfv\'en Wave Solar Model (AWSoM),
a global model from the
upper chromosphere to the corona and the heliosphere. The coronal heating and
solar wind acceleration are addressed with low-frequency Alfv\'en wave
turbulence. The injection of Alfv\'en wave energy at the inner boundary is
such that the Poynting flux is proportional to the magnetic field strength.
The three-dimensional magnetic field topology is simulated using data from
photospheric magnetic field measurements. This model does not impose
open-closed magnetic field boundaries; those develop self-consistently.
The physics includes:
(1) The model employs three different temperatures,
namely the isotropic electron temperature and the parallel and perpendicular
ion temperatures. The firehose, mirror, and ion-cyclotron instabilities
due to the developing ion temperature anisotropy are accounted for.
(2) The Alfv\'en waves are
partially reflected by the Alfv\'en speed gradient and the vorticity along the
field lines. The resulting counter-propagating waves are responsible
for the nonlinear turbulent cascade. The balanced turbulence due to
uncorrelated waves near the apex of the closed field lines and the resulting
elevated temperatures are addressed.
(3) To apportion the wave dissipation to the three temperatures, we employ the
results of the theories of linear wave damping and nonlinear stochastic
heating. (4) We have incorporated the collisional and collisionless
electron heat conduction.
We compare the simulated multi-wavelength EUV images of CR2107 with the
observations from {\it STEREO}/EUVI and {\it SDO}/AIA instruments. We
demonstrate that the reflection due to strong magnetic fields in proximity of
active regions intensifies the dissipation and observable emission
sufficiently.

\end{abstract}


\section{INTRODUCTION}

During the last few decades, considerable progress has been made in the
understanding of the solar atmosphere due to the increased availability of
observational data and the development of analytical and numerical models of
the solar wind. One aspect of this development is the construction of
complex three-dimensional (3D) models. These models can be validated with
observations and further refined to improve the comparison. Important in
the further progress of these models is a better understanding of the
coronal heating and solar wind acceleration problem. Improvements in the
theories of the coronal heating scenarios may result in more realistic models
that are more reliable in predicting the solar wind conditions. Eventually,
this may further improve the numerical forecasting of space weather events.

Recent Hinode observations suggested that there is more than enough energy
in the chromospheric magnetic field fluctuations, which propagate away from
the Sun, to heat the solar corona and maintain the temperatures at $1\;$MK
\citep{depontieu2007}. With the {\it Solar Dynamics Observatory} ({\it SDO})
these waves were shown to be ubiquitously present in the transition region and
low corona \citep{mcintosh2011}. These observations suggest that it is
therefore appealing to develop a 3D solar corona and inner heliosphere model
that is based on Alfv\'en wave turbulence and check if such a theoretical model
reproduces emission features seen in the extreme ultraviolet (EUV) images.
In our previous corona model \citep{sokolov2013}, we demonstrated that
Alfv\'en wave turbulence can indeed capture the overall observable EUV
emission, with the exception of some of the details such as
the emission around active regions.

There is a long history of Alfv\'en wave turbulence in the solar wind that
dates back to the pioneering work of \citet{coleman1968}, who concluded that
turbulence is important in the solar wind near $1\;$AU based on {\it Mariner 2}
measurements. The earliest solar wind models that
incorporated Alfv\'en wave turbulence were presented in \citet{belcher1971}
and \citet{alazraki1971}, followed by 2D global corona models by
\citet{usmanov2000} and \citet{hu2003}. \citet{suzuki2006} constructed
a self-consistent 1D Alfv\'en wave turbulence model from the photosphere to
inner heliosphere that included wave reflection and mode conversion from
Alfv\'en to slow waves. This model was generalized to 2D by
\citet{matsumoto2012} to account for turbulent cascade.
Alfv\'en waves that propagate outward from the Sun are partially reflected by
gradients due to stratification, which produces waves propagating in
opposite directions, see for instance \citep{heinemann1980,leroy1980,
matthaeus1999,dmitruk2002,verdini2007,cranmer2010,chandran2011}.
Counter-propagating waves are essential for the classical incompressible
cascade \citep{velli1989} and hence the coronal heating.
The reflection due to strong magnetic field
gradients may be important in close proximity of the active regions,
further intensifying the dissipation and the observable EUV emission.
In the present paper, we will determine if this enhanced wave reflection
does improve our model comparison with the observed EUV images.

Remote observations from the Ultraviolet Coronagraph Spectrometer (UVCS)
have shown that the perpendicular ion temperature is much larger than
the parallel ion temperature in the corona holes \citep{kohl1998,li1998}.
Similarly, Helios observations have shown that the ion temperature
in the inner heliosphere is anisotropic as well \citep{marsch1982}.
Several models have been constructed that take the ion temperature
anisotropy into account, see for instance \citet{leer1972,isenberg1984,
fichtner1991,chandran2011} for 1D analysis and \citet{vasquez2003,li2004}
for 2D analysis. The ion temperature anisotropy has, to the best of our
knowledge, not yet been incorporated in a 3D solar wind model.

In recent years, our solar wind efforts in the Space Weather Modeling Framework
(SWMF, \citet{toth2012}) were focused on creating an Alfv\'en wave
turbulence-driven solar corona and inner heliosphere model with a
two-temperature approach for the electrons and ions. Our goal is to
develop a single validated model that can produce realistic synthesized
line-of-sight (LOS) images in the multi-wavelength EUV, and accurate
$1\;$AU prediction of the solar wind properties. This model will also produce
realistic background solar wind for simulations of coronal mass ejections
(CMEs).
\citet{vanderholst2010} constructed a solar wind model that has an inner
boundary in the $1\;$MK corona and has separate electron and ion
temperatures. It incorporated the collisional electron heat conduction and
Alfv\'en wave transport and dissipation along the open
field lines. This model was validated in \citet{jin2012}, while
\citet{evans2012} enhanced the model by including surface Alfv\'en waves.
The two-temperature approach is important in producing a correct CME shock,
otherwise a strong and unphysical heat precursor can appear ahead of the CME
due to the heat conduction \citep{manchester2012,jin2013}. Meanwhile, 
\citet{downs2010} developed a lower corona model that was able to reproduce
synthesized EUV images, but for the coronal heating the model did still
rely on ad hoc coronal heating functions. \citet{sokolov2013} combined the
aforementioned efforts into a single two-temperature model with
Alfv\'en wave turbulence by incorporating the concept of balanced turbulence
at the apex of the closed field lines, and via a concise analysis, the model
was able to numerically resolve the upper chromosphere and transition region.
This model was able to reproduce the overall morphology of coronal holes and
active regions in the EUV images. \citet{landi2013b} demonstrated
that it was also able to capture the charge state evolution, while
\citet{oran2013} compared the model output to in-situ Ulysses observations.

\citet{lionello2009} demonstrated that a 3D magnetohydrodynamic (MHD) lower
corona model based on a combination of ad hoc coronal heating functions
and with heat conduction was
able to well reproduce many features in the observed EUV images.
\citet{odstrcil2005} showed that the ENLIL inner heliosphere MHD model
prescribed by the empirical Wang--Sheeley--Arge (WSA) model \citep{arge2000}
was able to capture the $1\;$AU observations, while
\citet{cohen2007,feng2010,vanderholst2010} used the WSA to prescribe the
coupled corona and inner inner heliosphere simulations to obtain $1\;$AU
results.

In the approach taken in the Alfv\'en wave solar model (AWSoM)
described in the current paper, we no longer rely on ad
hoc heating functions. The physics incorporated into the AWSoM model allows
us to produce realistic LOS images and $1\;$AU solar wind predictions in
one single model. We present the theoretical approach
and results in two companion papers. The present paper is the first
one and describes the coronal heating methodology and electron heat conduction
and presents the resulting synthesized images of the multi-wavelength EUV
emission. We also demonstrate the newly implemented low dissipation MP5 limiter
\citep{suresh1997} in the BATS-R-US model. The reduced numerical dissipation
allows us to clearly resolve the fine details in the solution, and allow a
better comparison with observations.
Our second paper, X. Meng et al. (in preparation), describes the impact of
these changes on the solar wind acceleration and the $1\;$AU comparisons.

The outline of the paper is as follows: In Section \ref{sec:model}, we
present details of the theoretical model and discuss the numerical
implementation. We first describe in Section \ref{sec:twotemperature}
the Alfv\'en wave turbulence approach for the two-temperature model.
In Section \ref{sec:turbulence} we provide the derivation of the Alfv\'en wave
propagation, reflection, and dissipation.
In Section \ref{sec:aniso}, we generalize this model
to three temperatures by using anisotropic ion temperatures and the isotropic
electron temperature. The implementation is described in Section
\ref{sec:setup}. In Section \ref{sec:results}, we demonstrate the performance
of this model for Carrington rotation (CR) 2107 by comparing EUV images with
observations. We conclude in Section \ref{sec:conclusions} and
provide details of the derivations in the appendices.

\section{COMPUTATIONAL MODEL}
\label{sec:model}

We now present the equations of the AWSoM model. We first describe the
two-temperature MHD equations for the electrons and ions to demonstrate
the incorporation of the Alfv\'en wave turbulence and collisionless heat
conduction. The details of the derivation of the Alfv\'en wave equations
are presented in Section \ref{sec:turbulence}.
In Section \ref{sec:aniso}, these equations are generalized to
ion temperature anisotropy. In Section \ref{sec:setup}, we give some details
of the numerical implementation. We finalize this section with a discussion on
the boundary conditions (Section \ref{sec:bc}). 

\subsection{Governing Equations of Two-temperature Model}
\label{sec:twotemperature}

The starting point of our model is the MHD equations, which well describe the
large-scale, low-frequency phenomena of the solar corona and inner heliosphere
plasma. In the inertial frame, the mass conservation, momentum conservation,
and induction equation are
\begin{equation}\label{eq:cont}
\frac{\partial\rho}{\partial t}+\nabla\cdot(\rho{\bf u})=0,
\end{equation}
\begin{equation}\label{eq:momentum}
\frac{\partial(\rho{\bf u})}{\partial t}+\nabla\cdot\left(\rho{\bf
  u}{\bf u}-\frac{{\bf B}{\bf
    B}}{\mu_0}\right)+\nabla\left(P_i+P_e+\frac{B^2}{2\mu_0}+P_A\right)=
-\rho\frac{GM_\sun}{r^3}{\bf r},
\end{equation}
\begin{equation}\label{eq:induction}
\frac{\partial {\bf B}}{\partial t}- \nabla\times({\bf u}\times{\bf B})=0,
\end{equation}
respectively. In addition, the initial conditions should satisfy
$\nabla\cdot{\bf B} = 0$.
The notation in these equations is as follows: $\rho$ is the mass density,
${\bf u}$ is the velocity, assumed to be the same for the ions and
electrons, ${\bf B}$ is the magnetic field, $G$ is the gravitational constant,
$M_\odot$ is the solar mass, $r$ is the position vector relative to the center
of the Sun, and $\mu_0$ is the permeability of vacuum.
The Alfv\'en wave pressure, $P_A$, provides additional
solar wind acceleration. The isotropic ion pressure $P_i$ and electron
pressure $P_e$ are determined by the energy equations:
\begin{eqnarray}\label{eq:energy}
\frac{\partial}{\partial
  t}\left(\frac{P_i}{\gamma-1}+\frac{\rho u^2}2+\frac{{\bf B}^2}{2\mu_0}\right)+\nabla\cdot\left[\left(\frac{\rho u^2}2+\frac{\gamma P_i}{\gamma-1}+\frac{B^2}{\mu_0}\right){\bf
  u}-\frac{{\bf B}({\bf u}\cdot{\bf B})}{\mu_0}\right] \nonumber\\
= -({\bf u}\cdot\nabla)\left(P_e+P_A\right)+
\frac{N_ik_B}{\tau_{ei}}\left(T_e-T_i\right)+Q_i-\rho\frac{GM_\sun}{r^3}{\bf
    r}\cdot{\bf u},
\end{eqnarray}
\begin{equation}\label{eq:electron}
\frac{\partial}{\partial t}\left(\frac{P_e}{\gamma-1}\right)
+\nabla\cdot\left(\frac{P_e}{\gamma-1}{\bf
  u}\right)+P_e\nabla\cdot{\bf u}=
  -\nabla\cdot{\bf q}_e
  +\frac{N_ik_B}{\tau_{ei}}\left(T_i-T_e\right)-Q_{\rm rad}+ Q_e.
\end{equation}
in which $T_{e,i}$ are the electron and ion temperatures, $N_{e,i}$ are
the electron and ion number densities, and $k_B$ is the Boltzmann constant.
We use the simple equation of state $P_{e,i} = N_{e,i}k_BT_{e,i}$ and the
polytropic index is $\gamma=5/3$. We have the
following additional energy contributions in the plasma energy equations.
The optically thin radiative energy loss in the lower corona is given by
\begin{equation}
  Q_{\rm rad} = N_e N_h \Lambda(T_e),
\end{equation}
where $N_h$ is the hydrogen number density and
$\Lambda(T_e)$ is the radiative cooling curve taken from the CHIANTI
version 7.1 database \citep{landi2013a}. The Coulomb collisional energy
exchange between the ions and electrons depends on the
relaxation time $\tau_{ei}$. The electron heat flux ${\bf q}_e$ consists of two
contributions. We use both the collisional formulation of Spitzer:
\begin{equation}
  {\bf q}_{e,S} = -\kappa_e T_e^{5/2}{\bf b}{\bf b}\cdot\nabla T_e,
\end{equation}
where ${\bf b} = {\bf B}/B$ and
$\kappa_e \approx 9.2\times 10^{-12}\;{\rm W}\,{\rm m}^{-1}\,{\rm K}^{-7/2}$,
as well as the collisionless heat flux as suggested by \cite{hollweg1978}:
\begin{equation}
  {\bf q}_{e,H} = \frac32 \alpha p_e {\bf u},
\end{equation}
in which we assume $\alpha=1.05$. We smoothly transition between these
formulations
\begin{equation}\label{eq:heatflux}
  {\bf q}_e = f_S {\bf q}_{e,S} + (1 - f_S){\bf q}_{e,H}.
\end{equation}
Here, the fraction of Spitzer heat flux is defined as a function of $r$
\begin{equation}\label{eq:interpolation}
  f_S = \frac{1}{1 + (r/r_H)^2},
\end{equation}
where $r_H = 5R_\odot$ similar to \citet{chandran2011}.
The Spitzer form is in this way the dominant
heat flux contributor near the Sun where the density is high enough so that the
electron temperature length scale $L_T=T_e/\left|\nabla T_e\right|$ is much
larger than the collisional mean free path of the electrons (this is
not correct for part of the transition region, but we ignore that fact in
the present paper), while far away from the
Sun the collisionless heat flux is more significant. In Appendix
\ref{sec:collisionless}, we describe the numerical implementation of the
collisionless heat flux. The electron and ion heating functions
are denoted by $Q_e$ and $Q_i$, respectively. Their sum equals the total
tubulence dissipation per unit time and per unit volume. The
partitioning of the dissipation into the coronal heating of the electrons and
ions is obtained from results of linear wave theory and stochastic heating,
see \cite{chandran2011} for details and Appendix \ref{sec:partition} for a
brief summary. To determine the Alfv\'en wave pressure and total wave
dissipation, we additionally solve for the propagation, reflection and
dissipation of the wave energy densities, $w_\pm$, in which the $+$ sign 
is for waves propagating in the direction parallel to ${\bf B}$, while the
$-$ sign is for waves propagating antiparallel to ${\bf B}$. The
turbulence equations are derived in Section \ref{sec:turbulence}. Here, we
summarize the final expressions for the time evolution of $w_\pm$:
\begin{equation}\label{eq:alfvenwaves}
\frac{\partial w_\pm}{\partial t}+\nabla\cdot\left[({\bf u}\pm{\bf
    V}_A)w_\pm\right]+\frac{w_\pm}2(\nabla\cdot{\bf u})=
\mp{\cal R}\sqrt{w_-w_+}
-\Gamma_\pm w_\pm
\end{equation}
where ${\bf V}_A = {\bf B}/\sqrt{\mu_0\rho}$ is the Alfv\'en speed.
The third term on the left hand side of Equation (\ref{eq:alfvenwaves})
represents the energy reduction in an expanding flow due to work done by the
Alfv\'en wave pressure $P_A=(w_++w_-)/2$. The last
term in Equation (\ref{eq:alfvenwaves}) is the wave dissipation per unit time
and per unit volume. It is expressed in the form of the phenomenological
dissipation rate
\begin{equation}
  \Gamma_\pm=\frac{2}{L_\perp}\sqrt{\frac{w_\mp}\rho},
\end{equation}
which contains the transverse correlation length of the
Alfv\'en waves in the plane perpendicular to ${\bf B}$. Similar to
\cite{hollweg1986}, we use a simple scaling law $L_\perp \propto B^{-1/2}$
with the proportionality constant $L_\perp\sqrt{B}$ as adjustable input
parameter of the model.
Since $\Gamma_\pm$ depends on the returning wave $w_\mp$, it is essential to
include the partial reflection of the forward propagating wave $w_\pm$.
The first term on the right hand side of Equation (\ref{eq:alfvenwaves}) is the
new source term describing the conversion into oppositely propagating waves.
The signed reflection rate ${\cal R}$ in this term is derived as
\begin{eqnarray}
&& {\cal R} = \min\left[{\cal R}_{\rm imb},\max(\Gamma_\pm)\right]
\left\{\begin{array}{ccl}\left(1-2\sqrt{\frac{w_-}{w_+}}\right)&{\rm
  if}&4w_-\le w_+\\0&{\rm
  if}&1/4w_-<w_+<4w_-\\ \left(2\sqrt{\frac{w_+}{w_-}}-1\right)&{\rm if}&4w_+\le w_-\end{array}\right., \label{eq:balanced}\\
&& {\cal R}_{\rm imb} = \sqrt{\left[({\bf V}_A\cdot\nabla)\log
    V_A\right]^2 + \left({\bf b}\cdot[\nabla\times
    {\bf u}]\right)^2}. \label{eq:imbalanced}
\end{eqnarray}
The reflection rate consists of three parts: (1) For the strongly imbalanced
turbulence, where $\min(w_\pm)/\max(w_\pm) \ll 1$, and moderately imbalanced
turbulence on open field lines and at the bottom of closed field lines, the
wave reflection rate is
represented by ${\cal R}_{\rm imb}$. In this case, the reflection is due
to Alfv\'en speed gradients and vorticity along the field lines. (2) We
additionally assume that the reflection rate is smaller than the maximum
dissipation rate to limit the reflection rate in the transition region,
which is accomplished by the min function.
(3) The full expression of the reflection rate (\ref{eq:balanced}) includes a
correction to the right of the curly bracket when the oppositely
propagating waves are of equal wave energy density near the apex of the
closed field lines. This correction presumes that the waves originating from
the two foot points are uncorrelated and as a result, the reflection rate is
negligible.

\subsection{Propagation, Reflection and Dissipation of Alfv\'en Waves}
\label{sec:turbulence}

In this section, we derive the Alfv\'en wave energy density equations
in several steps. In Section \ref{sec:wkbalfven}, we first derive the
WKB wave transport equation in the absence of wave reflection and dissipation.
The non-linear cascade rate is discussed in Section \ref{sec:dissrate}.
In Section \ref{sec:reflection},
the wave reflection is presented and we arrive at the final expressions for
the oppositely propagating waves. The limit of strongly imbalanced
turbulence is further analyzed in Appendix \ref{sec:localdiss}.

\subsubsection{WKB-Equation for Alfv\'en Turbulent Wave Energy Densities}
\label{sec:wkbalfven}

In this section, we derive the equations governing the evolution of the wave
energy densities, $w_{\pm}$. Our starting point is reduced MHD, which
solves the non-conservative form of Equations (\ref{eq:momentum})
and (\ref{eq:induction}):
\begin{equation}\label{eq:momentumnc}
\frac{\partial{\bf u}}{\partial t}+({\bf u}\cdot\nabla){\bf u}+
\frac{\nabla B^2}{2\mu_0\rho}+\frac{\nabla(P_e+P_i)}\rho
= \frac{({\bf B}\cdot\nabla){\bf B}}{\mu_0\rho}-\frac{GM_\sun}{r^3}{\bf r},
\end{equation}
\begin{equation}\label{eq:inductionnc}
\frac{\partial {\bf B}}{\partial t}+({\bf u}\cdot\nabla){\bf B}+{\bf B}(\nabla\cdot{\bf u})=({\bf B}\cdot\nabla){\bf u},
\end{equation}
as well as the continuity equation (\ref{eq:cont}).
We represent the magnetic field and velocity vectors as sums of 
regular and turbulent parts, ${\bf u}=\tilde{\bf u}+\delta{\bf u}$
and ${\bf B}=\tilde{\bf B}+\delta{\bf B}$
(below tildes are omitted) and simplify the 
equations for turbulent amplitudes by assuming the incompressibility
conditions $\nabla\cdot\delta{\bf u}=0$ and ${\bf B}\cdot\delta{\bf B}=0$:
\begin{equation}\label{eq:deltau}
\frac{\partial\delta{\bf u}}{\partial t}+({\bf u}\cdot\nabla)\delta{\bf u}+(\delta{\bf u}\cdot\nabla)\delta{\bf u}+(\delta{\bf u}\cdot\nabla){\bf u}
=\frac{({\bf B}\cdot\nabla)\delta{\bf B}}{\mu_0\rho}+\frac{(\delta{\bf B}\cdot\nabla)\delta{\bf B}}{\mu_0\rho}+\frac{(\delta{\bf B}\cdot\nabla){\bf B}}{\mu_0\rho},
\end{equation}
\begin{equation}\label{eq:deltaB}
\frac{\partial \delta{\bf B}}{\partial t}+({\bf u}\cdot\nabla)\delta{\bf B}+
(\delta{\bf u}\cdot\nabla)\delta{\bf B}+(\delta{\bf u}\cdot\nabla){\bf B}+\delta{\bf B}(\nabla\cdot{\bf u})=({\bf B}\cdot\nabla)\delta{\bf u}+(\delta{\bf B}\cdot\nabla)\delta{\bf u}+(\delta{\bf B}\cdot\nabla){\bf u},
\end{equation}
\begin{equation}\label{eq:deltarho}
\frac{\partial \rho}{\partial t}+({\bf u}\cdot\nabla)\rho+(\delta{\bf u}\cdot\nabla)\rho
+\rho\nabla\cdot{\bf u}=0.
\end{equation}
The equations for the Els\"asser variables, 
${\bf z}_\pm =\delta{\bf u}\mp\delta{\bf B}/\sqrt{\mu_0\rho}$ are obtained as 
the sum Equation (\ref{eq:deltau})$\mp\frac1{\sqrt{\mu_0\rho}}\times$ Equation
(\ref{eq:deltaB})$\pm\frac{\delta{\bf B}}{\rho\sqrt{\mu_0\rho}}\times$ Equation
(\ref{eq:deltarho}):
\begin{equation}
\frac{d_\pm{\bf z}_\pm}{dt}
+{\bf z}_\mp\cdot\nabla{\bf u}\mp
\frac{{\bf z}_\mp\cdot\nabla{\bf B}}{\sqrt{\mu_0\rho}}-
\frac{{\bf z}_\pm-{\bf z}_\mp}{4\rho}
\frac{d_\mp\rho}{dt}
=0, \label{eq:elsasser1}
\end{equation}
where $\frac{d_\pm}{dt}=\frac\partial{\partial t}+({\bf u}\pm{\bf
  V}_A+{\bf z}_\mp)\cdot\nabla$ and ${\bf V}_A=\frac{\bf
  B}{\sqrt{\mu_0\rho}}$ (see analogous equation in \cite{velli1993,
  chandran2009a,chandran2009b}). Note, that for the regular plasma
velocity and magnetic field the non-conservative form of
Equations (\ref{eq:momentum}) and (\ref{eq:induction}) is fulfilled, the wave
pressure resulting from the turbulent magnetic field being:
\begin{equation}\label{eq:wavepreslong}
P_A^{\rm full}=\frac{(\delta{\bf
    B})^2}{2\mu_0}=\rho\left(\frac{{\bf z}_+^2}8+\frac{{\bf z}_-^2}8-\frac{{\bf
  z}_-\cdot{\bf z}_+}4\right).
\end{equation}
The equations for the wave energy densities, $w_\pm=\rho{\bf z}^2_\pm/4$,
is obtained by the sum $\rho{\bf z}_\pm/2\times$ Equation
(\ref{eq:elsasser1}) $+3{\bf z}^2_\pm/8\times$ Equation
(\ref{eq:deltarho}):
\begin{equation}\label{eq:wpmfull}
\frac{\partial w_\pm}{\partial t}+\nabla\cdot[({\bf u}\pm{\bf V}_A+{\bf z}_\mp)w_\pm]+\frac{w_\pm}2(\nabla\cdot{\bf u})+\frac\rho2{\bf z}_\pm\cdot[({\bf z}_\mp\cdot\nabla){\bf u}\mp\frac{({\bf z}_\mp\cdot\nabla){\bf B}}{\sqrt{\mu_0\rho}}]+\frac{{\bf z}_+\cdot{\bf z}_-}8\frac{d_\mp\rho}{dt}=0. 
\end{equation}
As a first approximation we set the oppositely propagating wave
${\bf z}_\mp=0$ in the equations for $w_\pm$ and obtain:
\begin{equation}
\frac{\partial w_\pm}{\partial t}+\nabla\cdot\left[({\bf u}\pm{\bf V}_A)w_\pm\right]+\frac{w_\pm}2(\nabla\cdot{\bf u})=0.
\end{equation}
This WKB equation describes Alfv\'en wave propagation along the magnetic field
lines (first two terms) and the wave energy reduction in the expanding plasma 
(the last term) because of the work done by the wave pressure, the
latter in the WKB-approximation can be represented as:
\begin{equation}\label{eq:wavepressshort}
P_A=\frac12(w_++w_-),
\end{equation} 
This approximation is valid, if the waves propagate only in one
direction and we neglect reflection in a gradually varying
medium, so that we have in this case ${\bf z}_+\ne0$ and ${\bf z}_-=0$
or vice versa. Alternatively, one can consider the oppositely propagating
waves originating from two footpoints of the closed magnetic field line,
by assuming that the sources of these waves are not correlated. In this
case the quadratic in ${\bf z}_\pm$ wave energy densities, $w_\pm$,
produced by each of the
sources do not vanish, while the product of two random and
non-correlated amplitudes may be assumed to have a zero average: 
$\left<{\bf z}_+\cdot{\bf z}_-\right>=0$   

\subsubsection{Dissipation Rate}
\label{sec:dissrate}

Within the more accurate approximation (but still within the
WKB-method), the non-linear term $\nabla\cdot({\bf z}_\mp w_\pm)$ 
results in the turbulent cascade and the wave energy 
dissipation (see, e.g. \cite{dmitruk2002,chandran2009a}).
The dissipation rate for the wave energy density, $w_+$, is 
controlled by the amplitude of the oppositely propagating wave,
$|{\bf z}_-| =2\sqrt{w_-/\rho}$, and the correlation length, $L_\perp$,
in the transverse (with respect to the magnetic field)
direction, because for the Alfv\'en wave, propagating along the magnetic
field, $\delta{\bf u}$ and $\delta{\bf B}$ 
are perpendicular to the magnetic field. Therefore, 
$\nabla\cdot({\bf z}_\mp w_\pm)\sim\frac{2}{L_\perp}\sqrt{\frac{w_\mp}\rho}w_\pm$ and the WKB equations 
with an account for non-linear dissipation read:
\begin{equation}\label{eq:wavewkb}
\frac{\partial w_\pm}{\partial t}+\nabla\cdot\left[({\bf u}\pm{\bf V}_A)w_\pm\right]+\frac{w_\pm}2(\nabla\cdot{\bf u})
=-\Gamma_\pm w_\pm,
\end{equation}
\begin{equation}\label{eq:dissrate}
\Gamma_\pm=\frac{2}{L_\perp}\sqrt{\frac{w_\mp}\rho}
\end{equation}
These equations work for the balanced turbulence ($w_+=w_-=w$).
For imbalanced turbulence, they properly reduce the dissipation rate for
the dominant wave by expressing this rate in terms of the amplitude of the
minor oppositely propagating wave. 

The system of Equations (\ref{eq:cont})--(\ref{eq:electron}) once completed
with Equations (\ref{eq:wavewkb})--(\ref{eq:dissrate}) is consistent as long
as the wave pressure as in Equation (\ref{eq:wavepressshort}) is used. The
source of the dissipated turbulence energy in
Equations (\ref{eq:energy})--(\ref{eq:electron}) is balanced with the energy
sink in Equation (\ref{eq:wavewkb}). The wave pressure is applied in Equation
(\ref{eq:momentum}) as the momentum source, while the work done by the wave
pressure is, accordingly,
taken into account in Equation (\ref{eq:wavewkb}). The model consistency is
explicitly pronounced in the fact that the sum of energy equations
(\ref{eq:energy}), (\ref{eq:electron}), and (\ref{eq:wavewkb}) has a form of
a conservation law:
\begin{eqnarray}
&&\frac{\partial}{\partial
  t}\left(\frac{P_i+P_e}{\gamma-1}+\frac{\rho u^2}2+\frac{B^2}{2\mu_0}+w_++w_-\right)+\nabla\cdot\left[\left(\frac{\rho u^2}2+\frac{\gamma (P_i+P_e)}{\gamma-1}+\frac{B^2}{\mu_0}\right){\bf
  u}-\frac{{\bf B}({\bf u}\cdot{\bf B})}{\mu_0}\right] \nonumber\\
&&+\nabla\cdot\left[\left(w_++w_-+P_A\right){\bf
  u}+\left(w_+-w_-\right){\bf V}_A\right]=-\nabla\cdot{\bf q}_e-Q_{\rm rad}
-\rho\frac{GM_\sun}{r^3}{\bf r}\cdot{\bf u}.\label{eq:etotal}
\end{eqnarray}
The non-conservative sources in the right hand side, which do not have a
divergence-like form, are due to the energy losses for
emission and for the work done against the gravitational force. The
conservative property of the governing equations is important, since it
ensures the consistency of the model. Below, we carefully keep
the conservative property of the model while including the wave reflection.  

\subsubsection{The Model of Reflection}
\label{sec:reflection}

Equation (\ref{eq:elsasser1}) for ${\bf z}_\pm$ demonstrates that even in
a linear approximation, the WKB approach dismisses the correlations between
inward and outward propagating waves. Indeed, in the equation
(\ref{eq:elsasser1}) for
${\bf z}_+$ there are source terms linearly proportional to ${\bf z}_-$ and
vise versa,
while in the WKB approximation these source terms describing the
conversion between the oppositely going waves are omitted.
The omitted correlations are important for turbulent cascade
\citep{tu1995,dmitruk2002}. We include them into the solar model as
follows. First, in the part of the Els\"asser Equation (\ref{eq:elsasser1})
that is responsible for the wave reflection, we use the continuity
equation (\ref{eq:deltarho}) to obtain
\begin{equation}
\frac1\rho\frac{d_\mp\rho}{dt} =
-\nabla\cdot{\bf u}\mp{\bf V}_A\cdot\nabla\log\rho
 \mp \frac{1}{\sqrt{\mu_0\rho}}\delta {\bf B}\cdot\nabla\log\rho.
\end{equation}
The third term on the right hand side is a nonlinear term, which
can lead to wave reflection due to very steep density gradients in
the direction of the wave amplitude $\delta{\bf B}$. This direction is under
the imcompressibility
condition, ${\bf B}\cdot\delta{\bf B}=0$, transverse to the magnetic field.
Such surface reflection can for instance arise at coronal loop boundaries
and streamer boundaries, ultimately resulting in enhanced heating near these
locations. In the present paper, we do not consider these effects, so that
$\frac1\rho\frac{d_\mp\rho}{dt}\approx-\nabla\cdot{\bf u}\mp{\bf V}_A\cdot\nabla\log\rho$.
Now, we apply this transformation in Equation (\ref{eq:wpmfull}) and repeat
the above procedure to check how Equation (\ref{eq:etotal}) is
modified. Second, if we admit a non-zero correlator $\left<{\bf
  z}_+\cdot{\bf z}_-\right>$, then we should use the full magnetic pressure
(\ref{eq:wavepreslong}) in the momentum equation. This point is also
clear from the observation, that the work done by the wave pressure in
Equation (\ref{eq:wpmfull}) now becomes $[(w_\pm/2)-\rho({\bf z}_-\cdot{\bf
    z}_+/8)]\nabla\cdot{\bf u}$, with their total being equal to
$P_A^{\rm full}\nabla\cdot{\bf u}$. One can find that in the equation analogous to
(\ref{eq:etotal}) the wave pressure is changed for $P_A^{\rm full}$ and
the non-divergent term breaking the energy conservation appears in the
left hand side as follows:
\begin{eqnarray}
&& \frac{\partial}{\partial
  t}\left(\frac{P_i+P_e}{\gamma-1}+\frac{\rho u^2}2+\frac{B^2}{2\mu_0}+w_++w_-\right)+\nabla\cdot\left[\left(\frac{\rho u^2}2+\frac{\gamma (P_i+P_e)}{\gamma-1}+\frac{B^2}{\mu_0}\right){\bf
  u}-\frac{{\bf B}({\bf u}\cdot{\bf B})}{\mu_0}\right] \nonumber\\
&&+\nabla\cdot\left[\left(w_++w_-+P_A^{\rm full}\right){\bf
  u}+\left(w_+-w_-\right){\bf V}_A\right]+Q_{\rm noncons}=
-\nabla\cdot{\bf q}_e-Q_{\rm rad}
-\rho\frac{GM_\sun}{r^3}{\bf r}\cdot{\bf u},\\
&&Q_{\rm noncons}= \frac\rho2\left[{\bf z}_+\cdot({\bf
    z}_-\cdot\nabla){\bf u}+{\bf z}_-\cdot({\bf
    z}_+\cdot\nabla){\bf u}+\frac{{\bf z}_-\cdot({\bf
    z}_+\cdot\nabla){\bf B}-{\bf z}_+\cdot({\bf
    z}_-\cdot\nabla){\bf B}}{\sqrt{\mu_0\rho}}\right].
\end{eqnarray}
Since the tensor ${\bf z}_-{\bf z}_++{\bf z}_+{\bf z}_-$
is symmetric, one can find that the non-conservative energy source
involves a contribution proportional to the symmetric part of
the deformation velocity tensor,
$\frac{\partial u_i}{\partial x^j}+\frac{\partial u_j}{\partial x^i}$
and another term proportional to $\nabla\times{\bf B}$.
The energy source can also be expressed as follows:
\begin{equation}
Q_{\rm noncons}=
\rho\delta{\bf u}\cdot(\delta{\bf
  u}\cdot\nabla){\bf u}-\frac1{\mu_0}\delta{\bf B}\cdot(\delta{\bf
  B}\cdot\nabla){\bf u}+\frac1{\mu_0}\delta{\bf B}\cdot(\delta{\bf
  u}\cdot\nabla){\bf B}-\frac1{\mu_0}\delta{\bf u}\cdot(\delta{\bf
  B}\cdot\nabla){\bf B}.
\end{equation}
The reason for this energy non-conservation is that the system of Alfv\'en
waves, which are assumed to be transversely polarized ($\delta{\bf
  z}_\pm\cdot{\bf B}=0$) and interact with each other (via
reflection) and with the moving plasma (via the turbulent heating and
wave pressure), is not closed. Indeed,
one can take the following linear combination: $\rho{\bf
  u}\cdot$Equation (\ref{eq:deltau})$+\frac{{\bf B}}{\mu_0}\cdot$Equation
(\ref{eq:deltaB})$+\rho\delta{\bf u}\cdot{\bf u}\times$Equation
(\ref{eq:deltarho})+ $\delta{\bf u}\cdot$Equation
(\ref{eq:momentumnc})$+\frac{\delta{\bf B}}{\mu_0}\cdot$Equation
(\ref{eq:inductionnc})+$\frac{u^2}2\times$[Equation
  (\ref{eq:deltarho})$-$Equation (\ref{eq:cont})] (the
latter subtraction is applied because
$\frac{u^2}2\times$Equation (\ref{eq:cont}) contributes to
Equation (\ref{eq:energy}) and should not be doublecounted), which gives:
\begin{eqnarray}
&&\frac\partial{\partial t}\left(\rho{\bf u}\cdot\delta{\bf
  u}+\frac{\delta{\bf B}\cdot{\bf
    B}}{\mu_0}\right)+\nabla\cdot\left[\delta{\bf u}\left(\frac{\rho
    u^2}2+\frac{B^2}{\mu_0}\right)+\left({\bf u}+\delta{\bf u}\right)\left(\rho\delta{\bf u}\cdot{\bf u}+2\frac{\delta{\bf B}\cdot{\bf B}}{\mu_0}\right)\right] \nonumber \\
&&-\frac1{\mu_0}\nabla\cdot\left[\delta{\bf B}\left({\bf u}\cdot{\bf
    B}\right)+\left({\bf B}+\delta{\bf B}\right)\left(\delta{\bf
    u}\cdot{\bf B}+\delta{\bf B}\cdot{\bf u}\right)
\right]-Q_{\rm noncons}=0.
\end{eqnarray}
Here, we did not assume that $\nabla\cdot\delta{\bf u}=0$ and $\delta{\bf
  B}\cdot{\bf B}=0$, as we did before, and we account for the term, $\nabla(\delta{\bf
  B}\cdot{\bf B})$, in Equation (\ref{eq:deltau}) and the term $({\bf
  B}+\delta{\bf B})\nabla\cdot\delta{\bf u}$ in Equation (\ref{eq:deltaB}).
We also accounted for the term
$\delta{\bf B}\cdot{\bf B}\nabla\cdot{\bf u}$, which we omitted while
deriving Equation (\ref{eq:wpmfull}) under the assumption of
$\nabla\cdot\delta{\bf u}=0$.
We omitted all terms related to the ion and electron pressures
and their variations. We obtain yet another non-closed conservation law, which governs the
change in energy related to the variation in the  longitudinal
magnetic field. The non-convervative source, $Q_{\rm noncons}$ appears
to come with the opposite sign to the newly derived equation, meaning
that this source describes the
energy non-linear conversion from transverse Alfv\'en waves to
the compressible  mode
with $\delta{\bf
  B}\cdot{\bf B}\ne0$, i.e. magnetosonic waves. We arrive at an
important conclusion: the model of Equation (\ref{eq:wpmfull}) allowing a
correlation between
  the oscillations in the oppositely propagating waves is not closed
  unless the model for compressible turbulence is included. As long as,
here, we do not include such model, we are to eliminate the turbulent energy sink
to nowhere and add the term, $-Q_{\rm noncons}/2$, to the left hand
side of each of Equation (\ref{eq:wpmfull}):
\begin{eqnarray}
&& \frac{\partial w_\pm}{\partial t}+\nabla\cdot\left[({\bf u}\pm{\bf
    V}_A)w_\pm\right]+\left(\frac{w_\pm}2-\frac{\rho{\bf z}_+\cdot{\bf
    z}_-}8\right)(\nabla\cdot{\bf u}) \nonumber \\
&& =\frac\rho4\left[
[{\bf z}_\pm\times{\bf z}_\mp]\cdot[\nabla\times{\bf
    u}]\pm\frac{{\rm Tr}[({\bf z}_\pm{\bf z}_\mp+{\bf z}_\mp{\bf z}_\pm)\cdot\nabla{\bf B}]}{\sqrt{\mu_0\rho}}\pm
({\bf z}_+\cdot{\bf z}_-)({\bf V}_A\cdot\nabla)\log\sqrt{\rho}\right]
  \nonumber\\
&& -\Gamma_\pm w_\pm, 
\end{eqnarray}
where we used an easy-to-derive identity, ${\rm Tr}\left[\left({\bf
    a}{\bf b}-{\bf b}{\bf a}\right)\cdot\nabla{\bf c}\right]=[{\bf
    b}\times{\bf a}]\cdot[\nabla\times{\bf c}]$, for any three vectors, ${\bf
    a},{\bf b},{\bf c}$.
We can further simplify this expression by assuming the
transverse polarization of the Alfv\'en waves: ${\bf z}_\pm\cdot{\bf
  B}=0$. For $\delta{\bf B}$ part of the Els\"asser variables, ${\bf
  z}_\pm=\delta{\bf u}\mp\delta{\bf B}/\sqrt{\mu_0\rho}$, the 
condition, $\delta{\bf B}\cdot{\bf B}=0$, had been already introduced above, while the requirement for
the velocity oscillation 
$\delta{\bf u}$ to be approximately perpendicular to the magnetic field line is a
consequence of the assumed incompressibility property of the
turbulence, $\nabla\cdot\delta{\bf u}=0$, together with the expectation that the
turbulent wave vectors are directed along the magnetic field, so that
to satisfy the incompressibility condition the velocity oscillations
should, rather, be transverse. For transverse waves in the
3-by-3 tensor $(\nabla{\bf B})$ as present in the expression
for the wave conversion rate, one can leave only 2-by-2 transverse
components in the plane perpendicular to the magnetic field,
$(\nabla{\bf B})_{\perp\perp}$. Only the symmetric part of this tensor
matters, which possesses two eigenvalues (for curl-free, hence, potential
magnetic field,
these eigenvalues can be expressed in terms of the curvature radii of the
equipotential surface). If we neglect the difference between two
eigenvalues, we can admit that the symmetric part of the tensor is
proportional to 2-by-2 unity tensor, the proportionality coefficient
can be found by observing that the trace of the 2-by-2 unity tensor equals
2, while ${\rm Tr}\left[(\nabla{\bf
    B})_{\perp\perp}\right]=\nabla_\perp\cdot{\bf
  B}_\perp=\nabla\cdot{\bf B}-{\bf B}\cdot\nabla\log|{\bf B}|$. Under
the specified assumptions, the equation accounting for the Alfv\'en
turbulent wave reflection, that is, the energy exchange between two
oppositely propagating waves, which reduces the amplitude of the
outgoing wave and amplifies the incoming wave, reads:
\begin{eqnarray}
&& \frac{\partial w_\pm}{\partial t}+\nabla\cdot\left[({\bf u}\pm{\bf
    V}_A)w_\pm\right]+\frac{w_\pm}2(\nabla\cdot{\bf u}) \nonumber \\
&& =\mp\frac\rho4\left\{[{\bf z}_-\times{\bf z}_+]\cdot[\nabla\times{\bf
    u}]+
({\bf z}_+\cdot{\bf z}_-)({\bf V}_A\cdot\nabla)\log V_A\right\} 
-\Gamma_\pm w_\pm. \label{eq:reflviazpm}
\end{eqnarray}
Note, that we omitted the contribution from the correlator
${\bf z}_-\cdot{\bf z}_+$ to the work done by the wave pressure
(see the modified multiplier at
$\nabla\cdot{\bf u}$). Accordingly, we do not use this
contribution to the wave pressure in the momentum equation and return
to Equation (\ref{eq:wavepressshort}). The reason for this is that the
correlator ${\bf z}_-\cdot{\bf z}_+$ is assumed to be non-zero only
when the waves of one direction dominate over the opposite ones. In
this case, the contribution from this correlator to the wave pressure
is negligible compared with that from the dominant wave: $|{\bf
  z}_-\cdot{\bf z}_+|\le\|{\bf z}_-\|\|{\bf z}_+\|\ll\max({\bf
  z}_\pm^2)$. On the other hand, this contribution may be a large loss term
in the wave energy equation for the minor wave.
The neglect can be justified only if the reflection
coefficient is limited in such way, that the maximum admissible value,
corresponding to the equality case in the following estimate:
\begin{equation}
\frac{\left|[{\bf z}_-\times{\bf z}_+]\cdot[\nabla\times{\bf
    u}]+
({\bf z}_+\cdot{\bf z}_-)({\bf V}_A\cdot\nabla)\log
V_A\right|}{|{\bf z}_-||{\bf z}_+|}\le
\sqrt{\left({\bf b}\cdot[\nabla\times{\bf
    u}]\right)^2+\left[({\bf V}_A\cdot\nabla)\log
    V_A\right]^2},
\end{equation}
can be only achieved if one of the waves dominates over the opposite,
so that $\min(w_\pm)/\max(w_\pm)\ll1$. In this case, the amplitude and
polarization of the minor wave are imposed by those for the dominant
wave, the sign of the correlator is governed by the requirement that, in
the course of reflection, the dominant wave should decrease, the
oppositely propagating ``reflected'' wave should grow. If the
oppositely propagating waves are comparable, they are assumed to be 
non-correlated. The following choice of the final expression for the
reflection coefficient satisfies all the listed requirements:    
\begin{equation}\label{eq:wkbrefl}
\frac{\partial w_\pm}{\partial t}+\nabla\cdot\left[({\bf u}\pm{\bf
    V}_A)w_\pm\right]+\frac{w_\pm}2(\nabla\cdot{\bf u})=
\mp{\cal R}\sqrt{w_-w_+}
-\Gamma_\pm w_\pm
\end{equation}
where
\begin{eqnarray}\label{eq:refl}
&& {\cal R} = {\cal R}_{\rm imb}
\left\{\begin{array}{ccl}\left(1-2\sqrt{\frac{w_-}{w_+}}\right)&{\rm
  if}&4w_-\le w_+\\0&{\rm
  if}&1/4w_-<w_+<4w_-\\ \left(2\sqrt{\frac{w_+}{w_-}}-1\right)&{\rm if}&4w_+\le w_-\end{array}\right., \\
&& {\cal R}_{\rm imb} = \sqrt{\left({\bf b}\cdot[\nabla\times
    {\bf u}]\right)^2+\left[({\bf V}_A\cdot\nabla)\log
    V_A\right]^2}.
\end{eqnarray}
The above considerations may be limited by the assumption of a
small reflection coefficient, which, probably, should be less than
the dissipation rate, above the other criteria. Therefore,
Equation (\ref{eq:refl}) may be bounded as follows:
\begin{equation}
{\cal R} = \min\left[{\cal R}_{\rm imb},\max(\Gamma_\pm)\right]
\left[\max\left(1-2\sqrt{\frac{w_-}{w_+}},0\right)-\max\left(1-2\sqrt{
    \frac{w_+}{w_-}},0\right)\right].
\end{equation}
By comparing Equations (\ref{eq:reflviazpm}) and (\ref{eq:wkbrefl}) we can find
the correlators of amplitudes of the counter-propagating waves:
\begin{eqnarray}
&&\frac\rho4\left({\bf z}_-\cdot{\bf z}_+\right)=\frac{({\bf V}_A\cdot\nabla)\log
    V_A}{\left({\bf b}\cdot[\nabla\times{\bf
    u}]\right)^2+\left[({\bf V}_A\cdot\nabla)\log
    V_A\right]^2}{\cal R}\sqrt{w_-w_+},\\
&&\frac\rho4{\bf b}\cdot\left[{\bf
    z}_-\times{\bf z_+}\right]=\frac{\left({\bf b}\cdot[\nabla\times{\bf
    u}]\right)}{\left({\bf b}\cdot[\nabla\times{\bf
    u}]\right)^2+\left[({\bf V}_A\cdot\nabla)\log
    V_A\right]^2}{\cal R}\sqrt{w_-w_+}.
\end{eqnarray}
In this way, the contribution from the first correlator to the wave pressure,
$P^{\rm full}_A$ could be accounted for, but we do not do this in the present
paper and use Equation (\ref{eq:wavepressshort}). Equations (\ref{eq:wkbrefl})
and (\ref{eq:refl})
describing the Alfv\'en wave propagation, reflection and dissipation
close the system of the MHD equation in a physically consistent way, as
long as the energy conservation law, Equation (\ref{eq:etotal}), is fulfilled. 

The reflection model used in \citet{matthaeus1999} is very similar to ours
due to the following common features: (1) the reflection turns to zero in
balanced turbulence; (2) the sign of effect is such that the reflection
reduces the amplitude of the dominant wave and enhances the
counter-propagating minor wave; (3) the magnitude of the reflection
coefficient is controlled by the gradient of the Alfv\'en speed. Our
account of the vorticity is also not quite new, since the effect
of sheared flow on the mode conversion in the solar atmosphere is
discussed by \citet{hollweg2012,hollweg2013}. The important distinction is
that, for the sake of the model consistency and energy conservation, we
ruled out the non-conservative sources from Equations (\ref{eq:wkbrefl}).  
Now, we can evaluate the neglected terms, $Q_{\rm noncons}/2$ and
$-\frac\rho8({\bf z}_-\cdot{\bf z}_+)(\nabla\cdot{\bf u})$,
which might be added to the right hand side of Equations (\ref{eq:wkbrefl})
with the opposite sign:
$\frac\rho8({\bf z}_-\cdot{\bf z}_+)(\nabla\cdot{\bf u})-\frac{Q_{\rm noncons}}2$. 
In the low steady-state solar corona, the
plasma moves mostly along the magnetic field lines: ${\bf u}=u{\bf B}/B$,
which allows us to express the transverse components
of the velocity derivatives in terms of those of the magnetic field:
$(\nabla{\bf u})_{\perp\perp}=\frac{u}{B}(\nabla{\bf B})_{\perp\perp}$.
If we neglect the contribution from $\nabla\times{\bf B}$ into the
non-conservative source and express approximately
$\nabla\cdot{\bf u}\approx-({\bf u}\cdot\nabla)\log\rho$ from the continuity
equation, the energy source due to the Alfv\'en wave interaction with the
compression mode reads:
\begin{equation}
\frac\rho8
({\bf z}_-\cdot{\bf z}_+)(\nabla\cdot{\bf u})-
\frac{Q_{\rm noncons}}2\approx 
\frac\rho4({\bf z}_-\cdot{\bf z}_+)({\bf
    u}\cdot\nabla)\left(-\log\sqrt{\rho}+\log B\right)=\frac\rho4({\bf z}_-\cdot{\bf
        z}_+)({\bf u}\cdot\nabla)\log V_A.
\end{equation}
Thus, we do not introduce the mode conversion term proportional to 
$({\bf u}\cdot\nabla)\log V_A$ into our model.
The reason for this omission is that including this compressible MHD
turbulence term would break the energy conservation.

\subsection{Generalization to Ion Temperature Anisotropy}\label{sec:aniso}

Due to observational evidence of ion temperature anisotropy in the lower
corona \citep{kohl1998,li1998} and in the inner heliosphere \citep{marsch1982},
we have generalized our solar wind model to anisotropic ion temperatures.
The implementation and global magnetosphere application of the anisotropic
ion pressure is presented in \citep{meng2012a,meng2012b}. Here we will use
the same implementation in the solar context.

The equation for the ion pressure (\ref{eq:energy}) is now decomposed into two
equations for both the ion pressure component perpendicular to the magnetic
field, $P_{i\perp}$, and the ion pressure component parallel to the
magnetic field $P_{i\parallel}$. However, for convenience, we solve for the
averaged ion pressure $P_i=(2P_{i\perp} + P_{i\parallel})/3$ instead of
$P_{i\perp}$. The ion pressures are determined by the equations
\begin{eqnarray}
  &&\frac{\partial}{\partial t}\left( \frac{P_i}{\gamma -1}
  + \frac{\rho u^2}{2} + \frac{B^2}{2\mu_0}\right)
  + \nabla\cdot\left[ \left(\frac{\rho u^2}{2} + \frac{P_i}{\gamma - 1}
    + \frac{B^2}{\mu_0}\right){\bf u} + {\bf P}_i\cdot{\bf u}
    - \frac{{\bf B}({\bf u}\cdot{\bf B})}{\mu_0} \right] \nonumber \\
  &&= - {\bf u}\cdot\nabla(P_e+P_A)
  + \frac{N_ik_B}{\tau_{ei}}(T_e-T_i) + Q_i - \rho\frac{GM_\odot}{r^3}
  {\bf r}\cdot{\bf u},
  \label{eq:protonpressure}\\
  &&\frac{\partial P_{i\parallel}}{\partial t}
    + \nabla\cdot(P_{i\parallel}{\bf u})
    + 2P_{i\parallel}{\bf b}\cdot(\nabla{\bf u})\cdot{\bf b}
    = \frac{\delta P_{i\parallel}}{\delta t}
    + (\gamma - 1)\frac{N_ik_B}{\tau_{ei}}(T_e -T_{i\parallel})
    + (\gamma - 1)Q_{i\parallel},
  \label{eq:parallelproton}
\end{eqnarray}
where $T_{i\parallel}$ is the parallel ion temperature obtained from the
equation of state $P_{i\parallel} = N_ik_BT_{i\parallel}$ and
${\bf P}_i = P_{i\perp} I + (P_{i\parallel} - P_{i\perp}){\bf b}{\bf b}$ is the
ion pressure tensor. The second term
on the right hand sides of Equations (\ref{eq:protonpressure}) and
(\ref{eq:parallelproton}) are the collisional energy exchanges with the
electrons. The third term on the right hand sides are the heating functions
$Q_i$ and $Q_{i\parallel}$ for the averaged ion and parallel ion pressure,
respectively. The sum of the electron and averaged ion heating functions,
$Q_e+Q_i$, is equal to the total turbulence dissipation per unit volume per
unit time, $\Gamma_+w_+ + \Gamma_-w_-$. The partitioning of the wave
dissipation into $Q_e$, $Q_i$, and $Q_{i\parallel}$ is described in Appendix
\ref{sec:partition}. The first term on the right hand side of Equation
(\ref{eq:parallelproton}), $\delta P_{i\parallel}/\delta t$, is for the
relaxation of the pressure anisotropy by the parallel
firehose, mirror, and ion-cyclotron instability constraints. If those
instability criteria are met, we reduce the pressure anisotropy so that the
plasma is stable again. Details about the stability formulation, implementation
and results are given in \citet{meng2012b}.
The anisotropic ion pressure also modifies the
momemtum equation (\ref{eq:momentum}):
\begin{equation}
  \frac{\partial\rho{\bf u}}{\partial t} + \nabla\cdot\left[
    \rho{\bf u}{\bf u} + (P_{i\parallel} - P_{i\perp}){\bf b}{\bf b}
    - \frac{1}{\mu_0}{\bf B}{\bf B}\right]
    + \nabla\left( P_{i\perp} + P_e + \frac{B^2}{2\mu_0} + P_A \right)
    = -\rho\frac{GM_\odot}{r^3}{\bf r},
  \label{eq:momentum3T}
\end{equation}
in which the second term on the left hand side contains a new contribution
due to pressure anisotropy.
We further assume that the anisotropic pressure does not significantly change
the Alfv\'en wave turbulence, and hence we use the turbulence as
formulated for isotropic temperatures in Section \ref{sec:twotemperature}.

\subsection{Model Implementation}\label{sec:setup}

In this section we present some details of the implementation of the improved
solar wind model. This model uses the numerical schemes of the BATS-R-US
MHD solver and the overarching Space Weather Modeling Framework (SWMF),
see \citet{toth2012} for a description of the SWMF and BATS-R-US tools.
The SWMF is a software framework for modeling various space physics domains
in a single coupled model. It has been used, besides space weather
applications for the coupled Sun-Earth system, for many planetary, moon and
comet applications as well as the outer heliosphere. 
It has recently been extended to applications of radiation
hydrodynamics in the context of laser-driven high-energy-density physics
\citep{vanderholst2011,vanderholst2013}. The new components of the SWMF,
presented in this paper, are the solar corona (SC) and inner heliosphere (IH).

The SC model uses a 3D spherical grid with the radial coordinate ranging from
$1\;R_\odot$ to $24\;R_\odot$. The grid is highly stretched towards the Sun
with smallest radial cell size $\Delta r = 10^{-3}\;R_\odot$ to numerically
resolve the steep density gradients in the upper chromosphere. We
artificially broaden the transition region similar to that as described in
\citet{sokolov2013,lionello2009} to be able to resolve this region. The grid
is block decomposed using the block-adaptive tree library (BATL, 
\citet{toth2012}). This library is a tool to create, load balance and message
pass the adaptive refined mesh and solution data. In the simulations of this
paper, the grid blocks consist of $6\times4\times4$ mesh cells. Inside
$r=1.7\;R_\odot$, the angular resolution is 256 cells in longitude and
128 cells in latitude corresponding to an angular cell size of $1.4^\circ$,
while outside that radius the grid is one level less
refined. The system of equations described in Sections \ref{sec:twotemperature}
and \ref{sec:aniso} are solved in the heliographic rotating frame by
including centrifugal and Coriolis forces
$-\rho\left[ {\bf \Omega}\times({\bf \Omega}\times
{\bf r}) + 2{\bf \Omega}\times{\bf u}\right]$ in the momentum equation and
adding the centrifugal contribution
$-\rho{\bf u}\cdot[{\bf \Omega}\times({\bf \Omega}\times{\bf r})]$
to the ion energy equations (\ref{eq:energy}) and (\ref{eq:protonpressure}).
Here ${\bf \Omega}$ is the angular velocity of
the Sun. We assume a uniform solar rotation with a 25.38 days period
so that $\Omega = 2.865\times10^{-6}\;{\rm rad}\,{\rm s}^{-1}$. For steady
state simulations, we use local time stepping, which speeds up the
convergence relative to time accurate simulations.
During the steady state convergence, we apply one additional level of mesh
refinement at the heliospheric current sheet.
To resolve the details in the LOS EUV images, we also demonstrate
higher resolution in latitude and longitude by using $6\times 6\times 6$
grid blocks and hence, an angular cell size of $0.94^\circ$ near the Sun,
in combination with the numerical scheme based on the spatially 5th-order MP5
limiter \citep{suresh1997} instead of our standard second-order shock-capturing
schemes \citep{toth2012}.

Details of the inner heliosphere setup and simulations are provided in
X. Meng et al. (in preparation).

\subsubsection{Boundary Conditions}\label{sec:bc}

Here, we limit our discussion to the pre-specified boundary conditions only,
and refer the reader to \citet{sokolov2013} for a more complete description.
The radial magnetic field component $B_r$ is
prescribed using synoptic magnetogram data in the following way: This data is
first extrapolated to a 3D potential field source surface (PFSS) solution
using either spherical harmonics or the finite difference iterative
potential-field solver (FDIPS). In the current paper, we use FDIPS since this
method avoids the ringing patterns near regions of concentrated
magnetic fields to which the spherical harmonics method is susceptible, see
\citet{toth2011}. The PFSS magnetic field is used both as the initial
condition and to set the boundary conditions.

The boundary condition for the Alfv\'en wave energy density is empirically set
by prescribing the Poynting flux of the outgoing waves ($w$ is $w_+$
for positive $B_r$ and $w_-$ for negative $B_r$): $S_A = V_Aw\propto B_\odot$,
where $B_\odot$ is the field strength at the inner boundary and the
proportionality constant is estimated in \citet{sokolov2013} as
$(S_A/B)_\odot = 1.1\times 10^6\;{\rm W}\,{\rm m}^{-2}\,{\rm T}^{-1}$.
Under the assumption of sufficiently small returning flux, this estimate of the
Poynting-flux-to-field ratio is equivalent to the following averaged
velocity perturbation
\begin{equation}
  <\delta{\bf u}_\perp\cdot\delta{\bf u}_\perp>^{1/2} \approx
  15\;{\rm km}\,{\rm s}^{-1}\left( \frac{3\cdot10^{-11}\;{\rm kg}\,
      {\rm m}^{-3}}{\rho}\right)^{1/4},
\end{equation}
where the mass density $3\cdot10^{-11}\;{\rm kg}\,{\rm m}^{-3}$ (ion number
density $N_i=2\cdot10^{16}\;{\rm m}^{-3}$) corresponds to the
upper chromosphere. This value is compatible with the Hinode
observations of the turbulent velocities of $15\,{\rm km}\,{\rm s}^{-1}$
\citep{depontieu2007}. Hence, the energy density of the outgoing wave is set to
$w = (S_A/B)_\odot \sqrt{\mu_0\rho}$. The returning wave energy density is
absorbed by setting it to zero.

The temperatures are all set to the same value
$T_e = T_i = T_{i\parallel} = T_\odot = 50,000\;{\rm K}$ uniformly at the inner
boundary. In \citet{sokolov2013} it was demonstrated that the grid spacing of
$\Delta r = 10^{-3}\;R_\odot$ is, in this case,
sufficient to numerically resolve the
density scale height. We overestimate the density for this temperature
by an order of magnitude
with the value of $N_e = N_i = N_\odot = 2\times 10^{17}\;{\rm m}^{-3}$ at the
inner boundary.
This overestimate prevents chromospheric evaporation and extends the upper
chromosphere to reach the correct lower density, but does not significantly
change the global solution as shown in \citet{lionello2009}.

\section{SIMULATION RESULTS FOR CR2107}
\label{sec:results}

For this paper, we selected CR2107 (2011 February 16 through March 16). This
rotation was also used in the validation studies of \citet{sokolov2013}.
In that paper, we were able to reproduce the overall morphology of the coronal
holes and active regions in the LOS EUV images of {\it SDO} and {\it STEREO}.
However, details in those images did not show up.
It is the goal of the present section to demonstrate that with the new version
of the AWSoM model, we are now able to produce high quality synthesized images
that capture details of the EUV observations. The validation of this
model with in situ data at $1\;$AU will be presented in
X. Meng et al. (in preparation).

\begin{figure}
\begin{center}
{\resizebox{0.48\textwidth}{!}{\includegraphics[clip=]{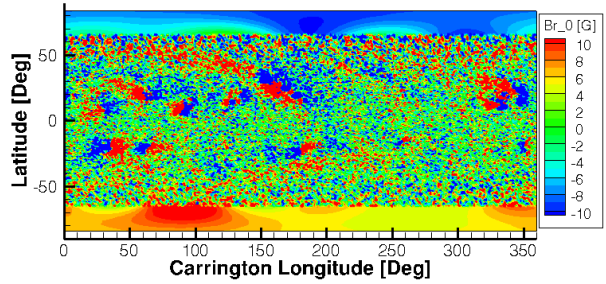}}}
\end{center}
\caption{Carrington map of the radial magnetic field component at
$1\;R_\odot$. This map is based on a synoptic magnetogram of CR2107
from {\it SDO}/HMI and processed to a PFSS solution using FDIPS.
For the purpose of showing both the active regions and coronal holes, we have
saturated the magnetic field in this plot at $\pm10\;$G.}
\label{fig:br}
\end{figure}

To simulate a background solar coronal solution, we need to specify
the radial magnetic field component at the inner boundary, which is located
in the upper chromosphere. We obtain this magnetic field in the following
way: The synoptic map CR2107 of the {\it SDO}/Helioseismic and Magnetic Imager
(HMI) is used. The polar field of this map is corrected with a
two-dimensional and third-order polynomial fitting of the data above $60^\circ$
\citep{sun2011}. We use FDIPS to generate an initial condition for the
magnetic field and the boundary condition values for the radial magnetic field
component (see Figure \ref{fig:br}). 

In all our results we will use parameter values
as summarized in Table \ref{table:parameters}, unless stated otherwise. The
boundary condition values are the same as in \citet{sokolov2013}. The value of
$L_\perp\sqrt{B}$ is twice larger compared to the value used in
\citet{sokolov2013} due to a factor two difference in the
definition of this parameter. The stochastic heating parameter $h_s$ and
the collisionless heat conduction parameter $r_H$ are assigned
with the same values as in \citet{chandran2011}.

\subsection{Heat Partitioning}

\begin{figure}
\begin{center}
{\resizebox{0.37\textwidth}{!}{\includegraphics[clip=]{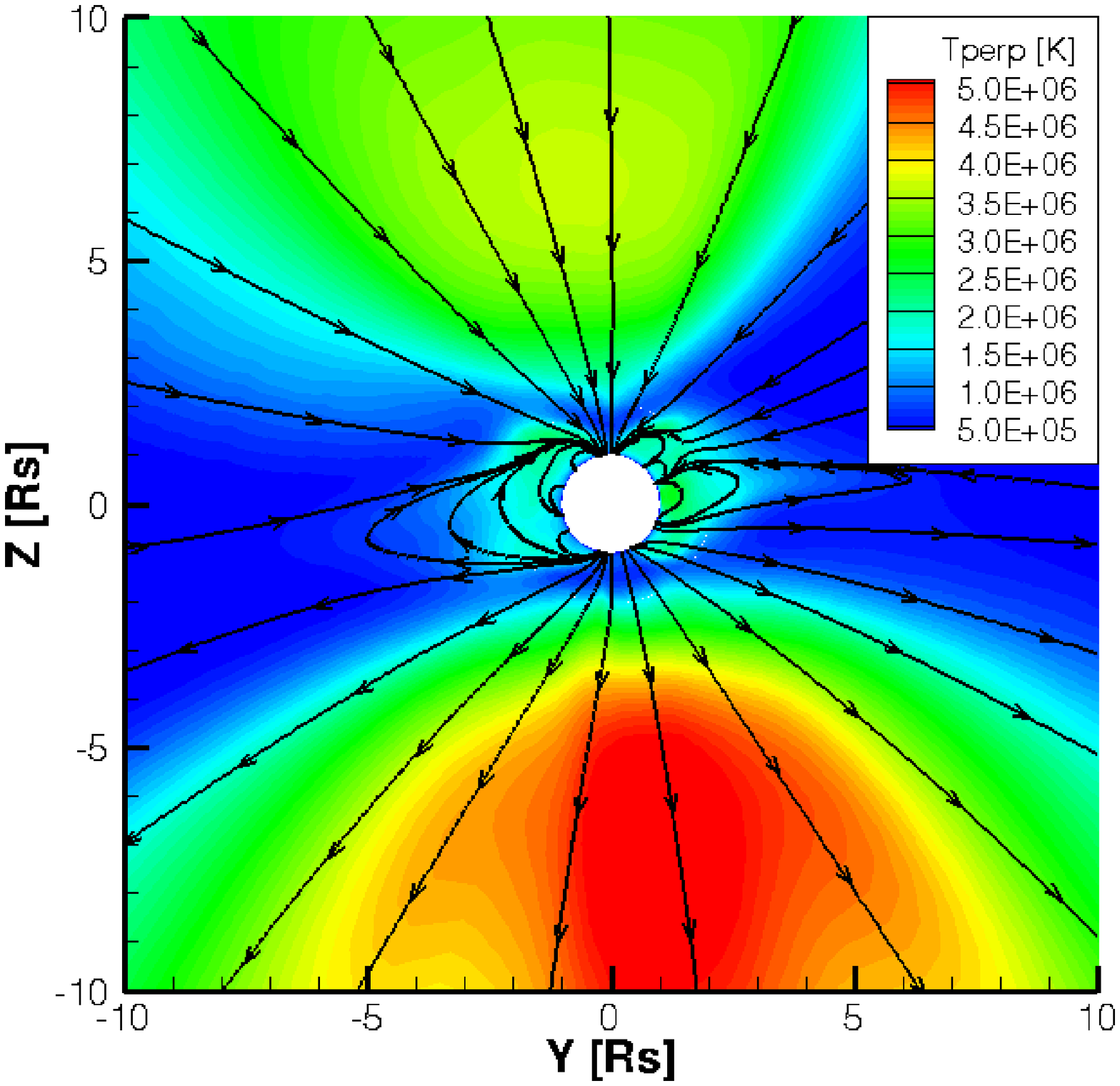}}}
{\resizebox{0.37\textwidth}{!}{\includegraphics[clip=]{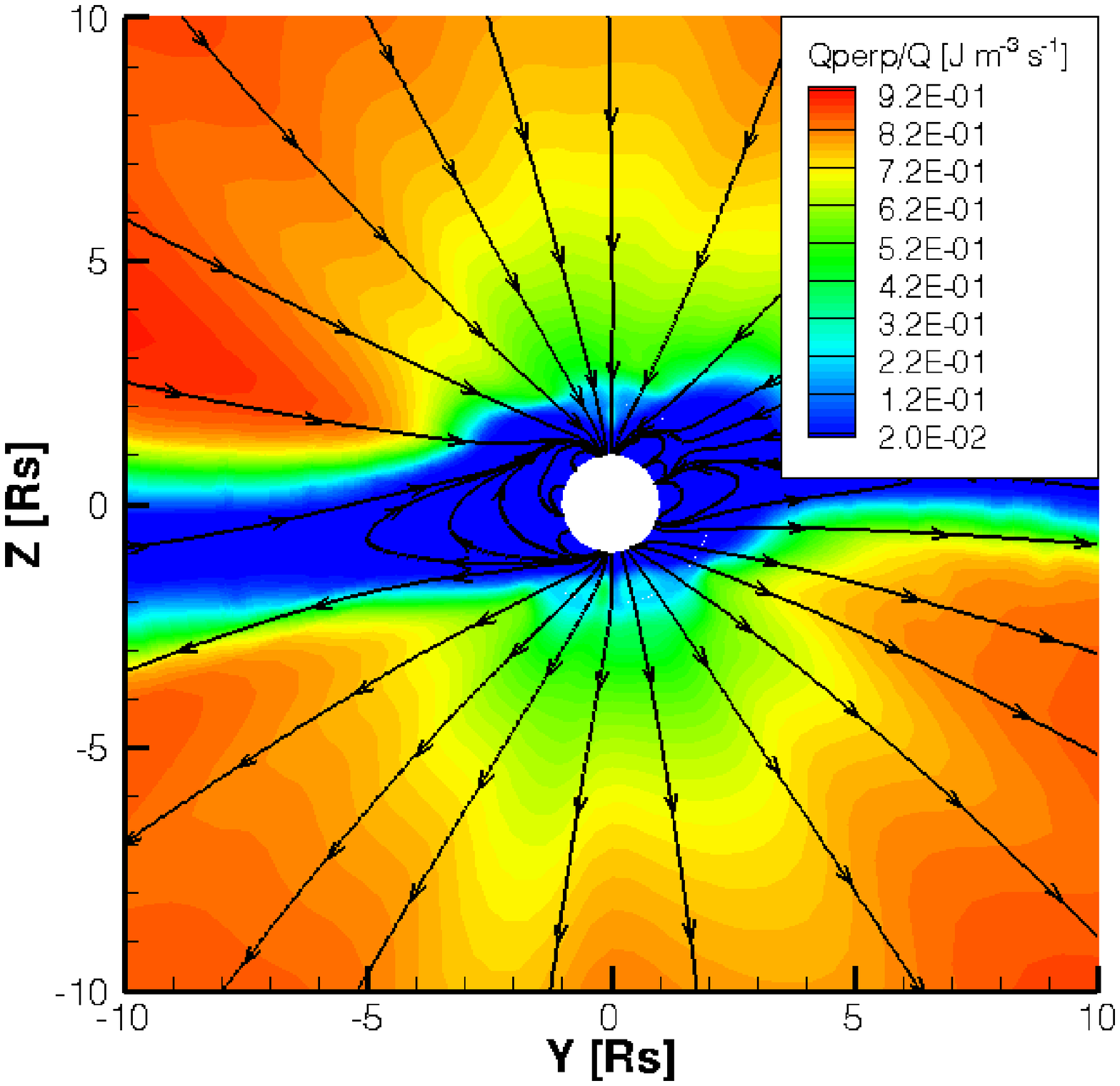}}}\\
{\resizebox{0.37\textwidth}{!}{\includegraphics[clip=]{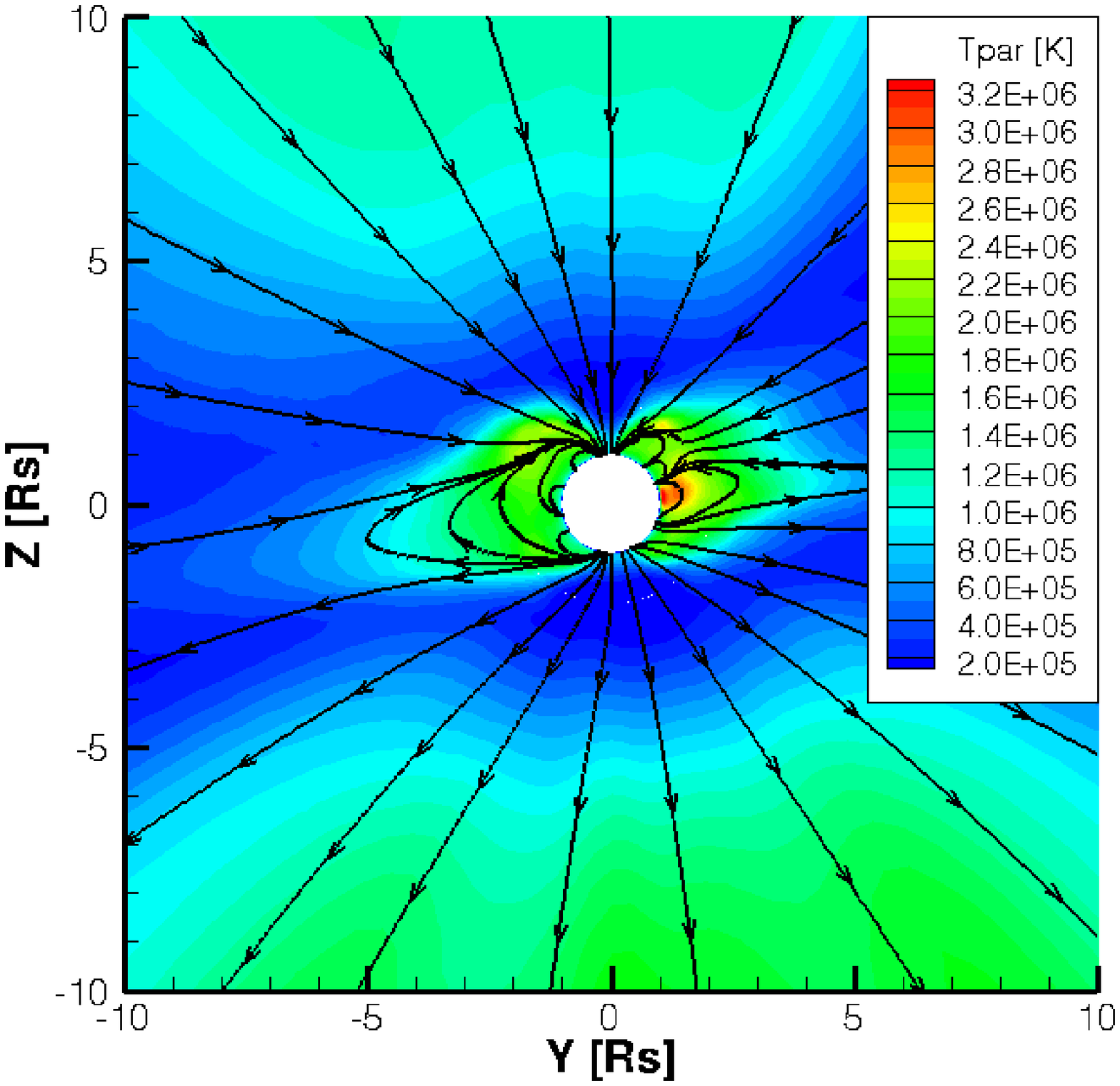}}}
{\resizebox{0.37\textwidth}{!}{\includegraphics[clip=]{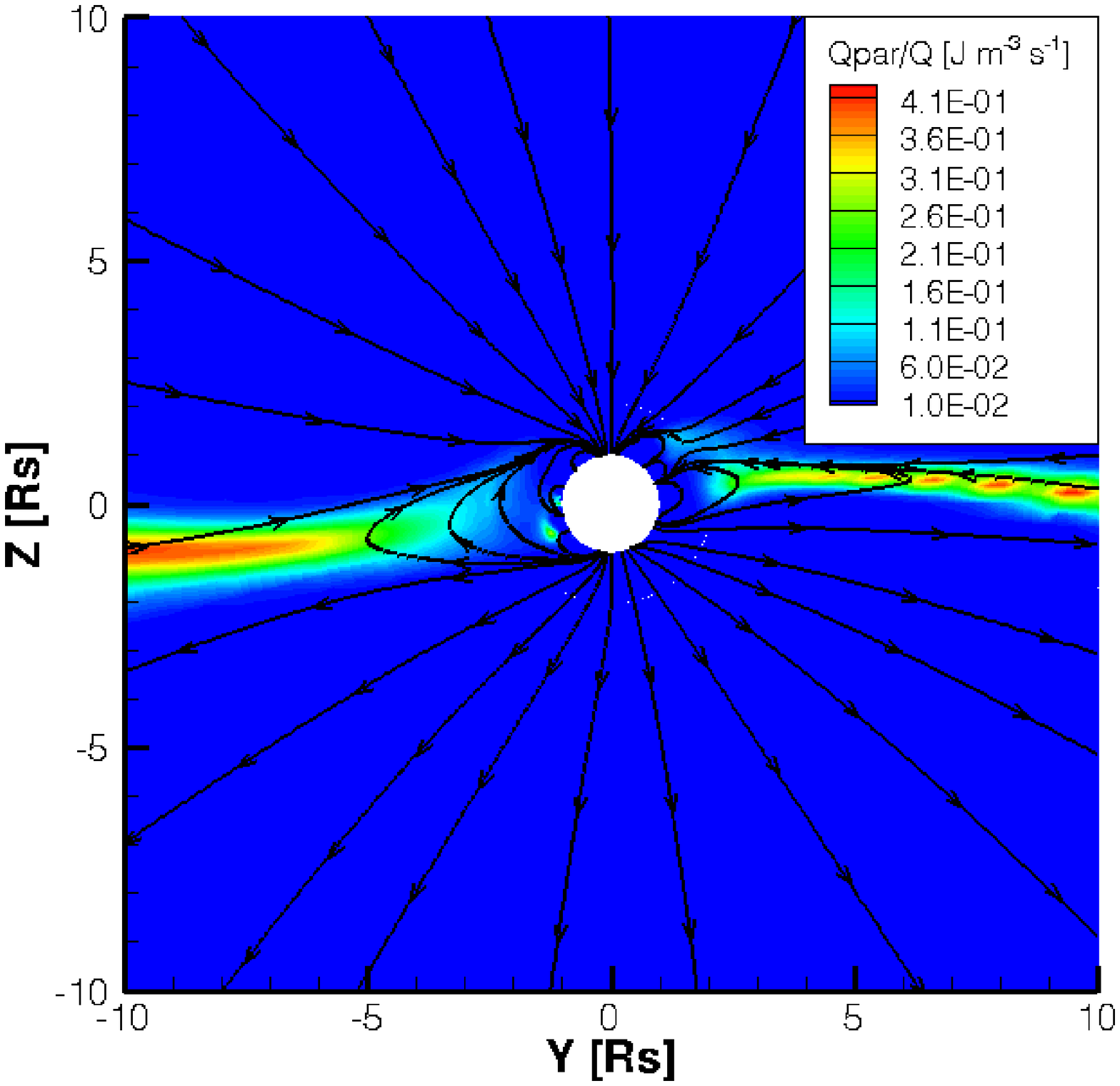}}}\\
{\resizebox{0.37\textwidth}{!}{\includegraphics[clip=]{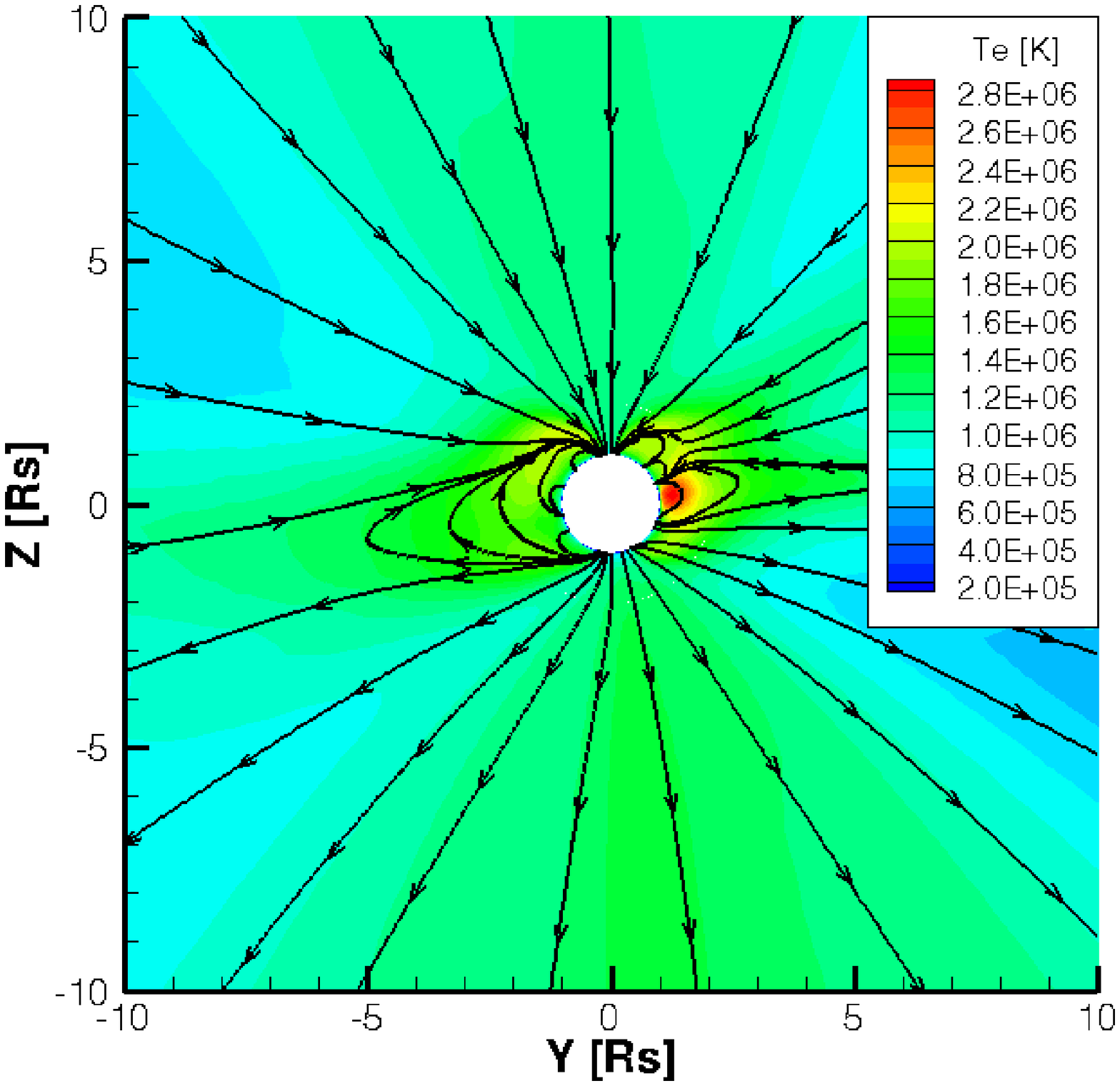}}}
{\resizebox{0.37\textwidth}{!}{\includegraphics[clip=]{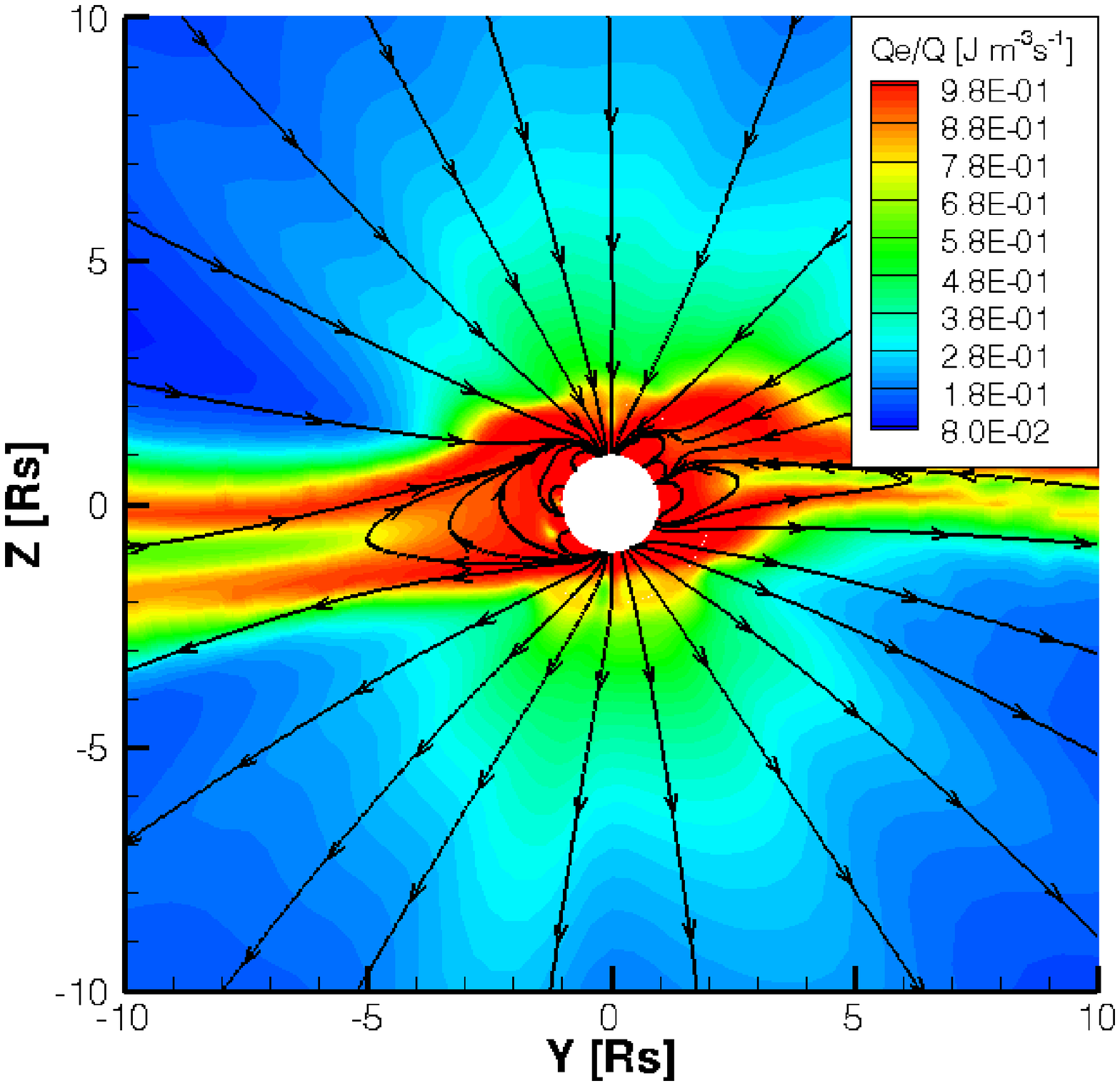}}}
\end{center}
\caption{Meridional slice ($X=0$ plane from $-10\;R_\odot$ to $10\;R_\odot$)
of the lower corona showing the three temperatures and heating fractions.
Left panels (from top to bottom): perpendicular ion temperature, parallel
ion temperature, and electron temperature in color contour, respectively.
Streamlines represent field lines by ignoring the out-of-plane component.
Right panels (from top to bottom): the ratio of the perpendicular ion,
parallel ion, and electron coronal heating with the total turbulence
dissipation.}
\label{fig:temperature}
\end{figure}

We will first demonstrate the heat partitioning of the
turbulence dissipation for our three-temperature model. In this simulation, we
used the version of this model with the wave reflection term in Equations
(\ref{eq:alfvenwaves})--(\ref{eq:imbalanced}). The steady state solution is
obtained with the spatially second-order shock-capturing scheme. In Figure
\ref{fig:temperature}, we show the three obtained temperatures $T_{i\perp}$,
$T_{i\parallel}$, and $T_e$ from top to bottom in the panels on the left.
These temperatures are shown in the meridional slice $X=0$ along with a few
projected field lines to indicate the location of open and closed field lines.
Very close to the Sun, the three temperatures are nearly the same. This is due
to the high density near the Sun, resulting in a sufficiently high
rate of Coulomb collisions that equilibrate the temperatures. The collision
rate decreases with the density, so that further away from the Sun
the collisions are too infrequent to equilibrate the temperatures.
The significant heating of the perpendicular ion temperature in the polar
coronal holes is due to the stochastic heating.
The heat partitioning fractions of the coronal heating
into ion perpendicular, ion parallel, and electron heating are shown in
the panels on the right from top to bottom, respectively. The ion
perpendicular heating, due to the stochastic heating process, dominates in the
lower corona sufficiently far away from both the Sun and the heliospheric
current sheet. The parallel ion heating is only
significant very close to the heliospheric current sheet (HCS) due to the
high plasma beta $\beta_i$, while the electron heating is important very
close to the Sun and around the (HCS).

\subsection{EUV Comparison}

From the density and electron temperature distribution, calculated with our
solar model, we can produce synthesized EUV that can
be compared with the observed ones. Such a comparison serves as a check for the
performance of the coronal heating model. The presented model accounts
for the partial reflection of the outward propagating waves, which is
accompanied by the generation of counter-propagating waves. These oppositely
propagating waves are ultimately responsible for the turbulent cascade rate
and hence, the coronal heating. The distinct feature in the present model is
the enhanced reflection in the presence of strong magnetic fields, such as in
close proximity of active regions, that can increase the dissipation and
thereby intensify the observable EUV emission.

To better resolve the details in
the synthesized LOS images, the latitudinal and longitudinal resolution is
increased by using adaptive mesh refinement grid size of $6\times6\times6$. 
The first attempt to use the full model as described in Section
\ref{sec:model} did not yet provide us the desired LOS image quality.
The problem is due to the numerical inaccuracy in the Alfv\'en speed
gradients in the reflection source term. To overcome this issue, we plan to
solve the upper chromosphere and transition region semi-analytically in a
forthcoming paper. In the present
paper we changed for now to the turbulence model that is based on local
dissipation in Equations (\ref{eq:wkblocal}) and (\ref{eq:disslocal}) instead
of the turbulence model with the wave reflection term in Equations
(\ref{eq:alfvenwaves})--(\ref{eq:imbalanced}). This model is less
susceptible by these numerical errors.

\begin{figure}
\begin{center}
{\resizebox{0.86\textwidth}{!}{\includegraphics[clip=]{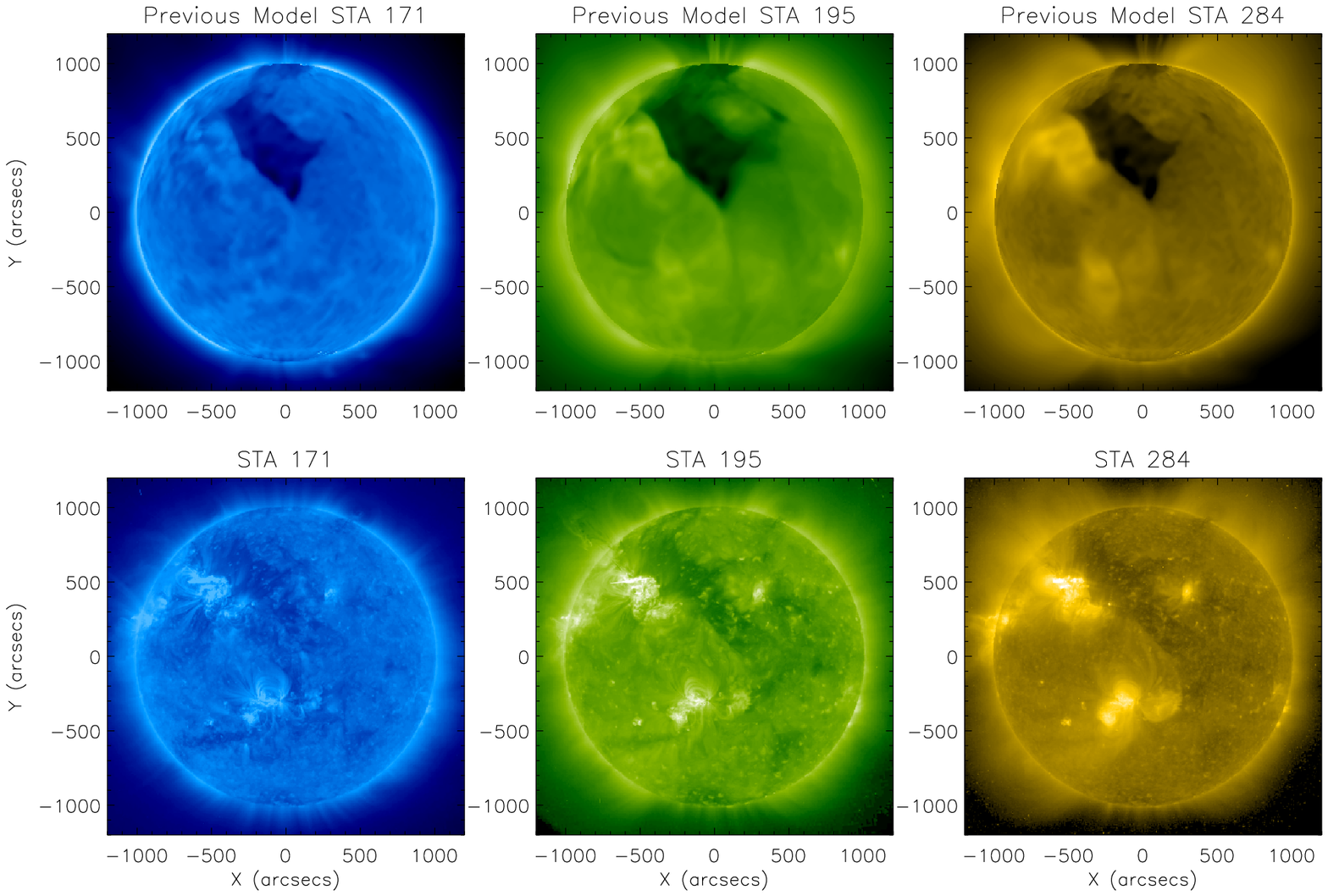}}}
{\resizebox{0.86\textwidth}{!}{\includegraphics[clip=]{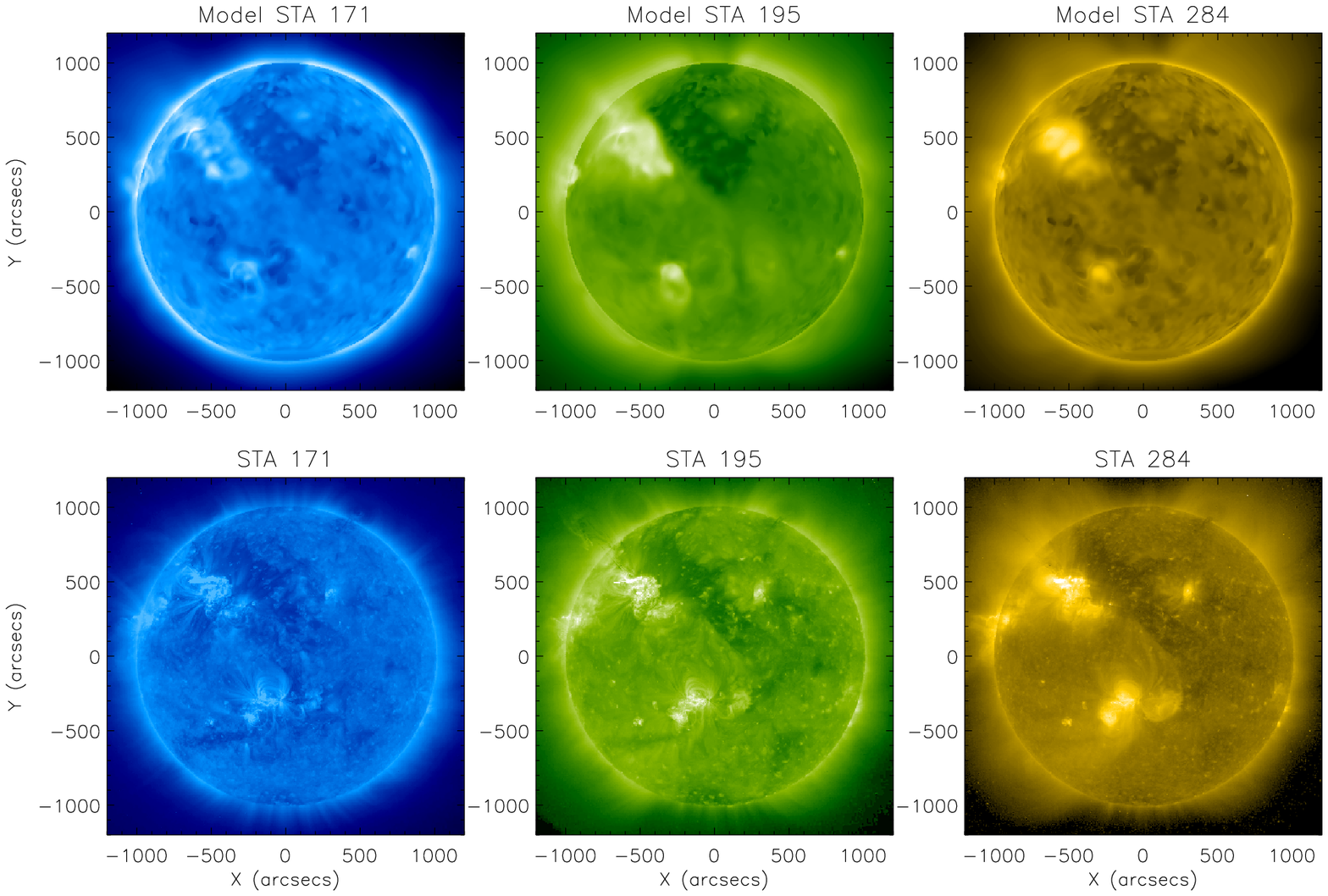}}}
\end{center}
\caption{Comparison of synthesized EUV images of the model with observational
{\it STEREO A}/EUVI images. The columns are from left to right for
$171\;{\rm \AA}$, $195\;{\rm \AA}$, and $284\;{\rm \AA}$. Top panels:
synthesized EUV images of the \citet{sokolov2013} model.
Midlle panels: synthesized EUV images of the improved model.
Bottom panels: observational {\it STEREO A}/EUVI images.
The observation time is 2011 March 7 20:00$\;$UT.}
\label{fig:stereoa}
\end{figure}

\begin{figure}
\begin{center}
{\resizebox{0.86\textwidth}{!}{\includegraphics[clip=]{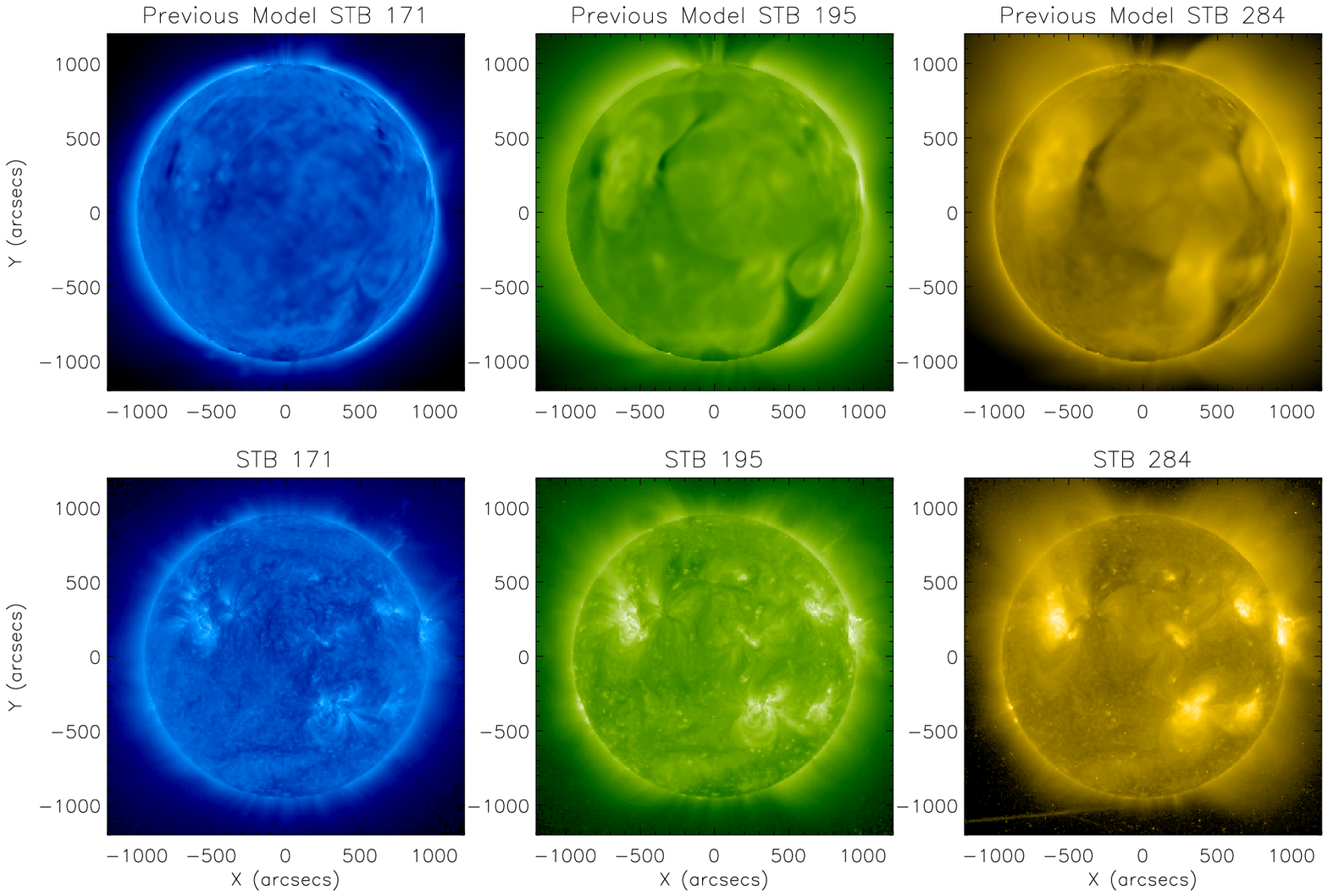}}}
{\resizebox{0.86\textwidth}{!}{\includegraphics[clip=]{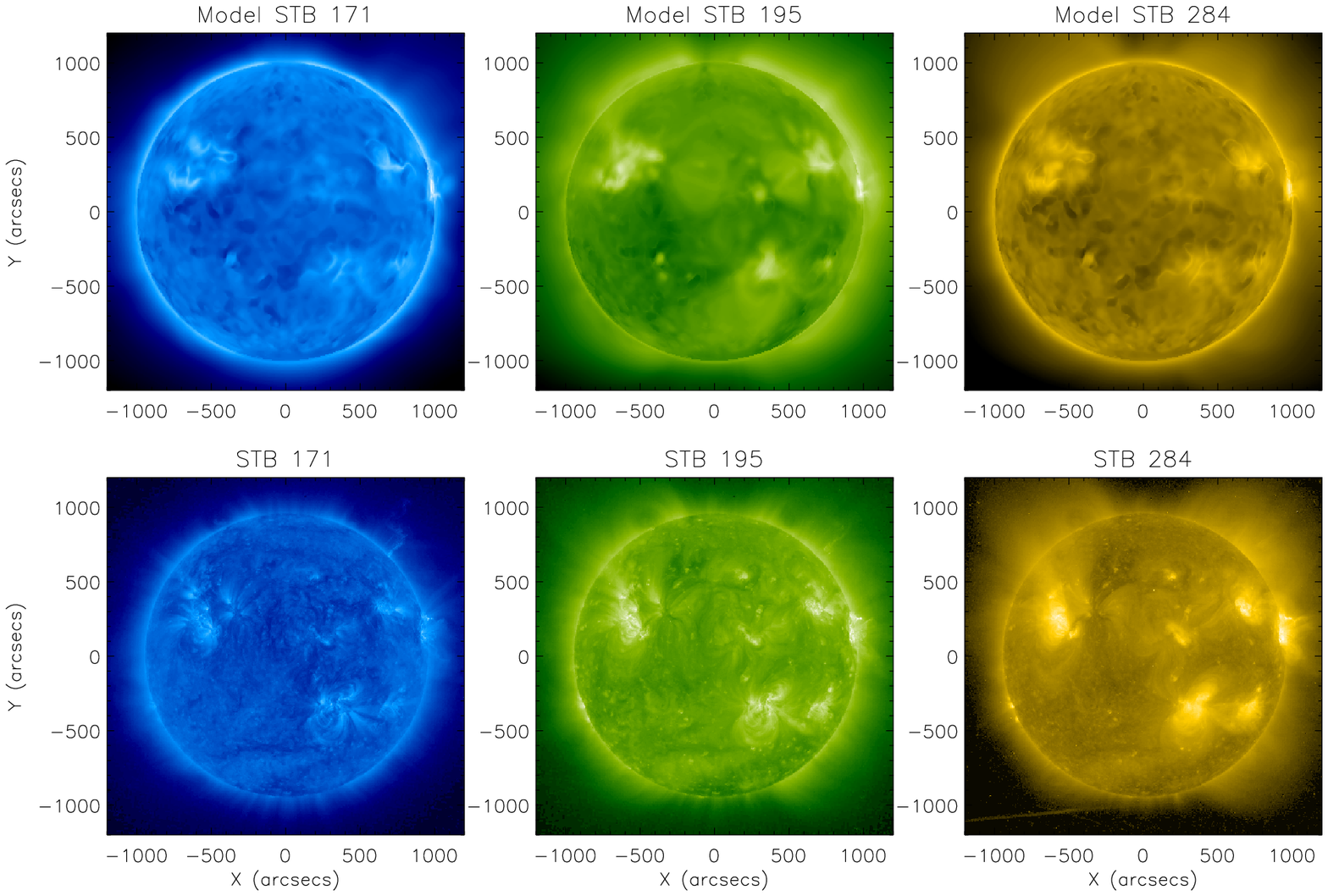}}}
\end{center}
\caption{Comparison of synthesized EUV images of the model with observational
{\it STEREO B}/EUVI images. The columns are from left to right for
$171\;{\rm \AA}$, $195\;{\rm \AA}$, and $284\;{\rm \AA}$. Top panels:
synthesized EUV images of the \citet{sokolov2013} model.
Midlle panels: synthesized EUV images of the improved model.
Bottom panels: observational {\it STEREO B}/EUVI images.
The observation time is 2011 March 7 20:00$\;$UT.}
\label{fig:stereob}
\end{figure}

In Figure \ref{fig:stereoa}, we computed the {\it STEREO}/EUVI emission
images in the three coronal bandpasses for Fe emission lines at 171, 195, and
$284\;$\AA. These LOS images are produced by assuming that the plasma is
optically thin for all the considered wavelengths. In the top row, the images
are for the CR2107 steady solution of our previous model \citep{sokolov2013}
using the spatially fifth-order MP5 limiter. In the middle row, we
demonstrate the new model with the MP5 limiter. The new model better
captures the active region emissivity as observed by the EUVI imager
\citep{howard2008} on board {\it STEREO A}, as shown in the bottom panels.
This enhanced emissivity is due to the increased reflection rate caused by
the strong magnetic fields around active regions.
We note that the steady state simulation was performed for a
synoptic magnetogram, while the observation is for the time 2011 March 7
20:00$\;$UT, and consequently, the model can not reproduce time dependent
activity during the rotation. Also the polynomial extrapolation towards the
pole in the CR2107 magnetogram might distort the high latitudinal region
somewhat unfavorably. The observed polar coronal holes are somewhat wider than
the coronal holes of the new model. In Figure \ref{fig:stereob}, we similarly
plot the results for {\it STEREO B}, which shows the other side of the Sun
as the two {\it STEREO} spacecraft are separated by about 177 degrees. The
emissivity of the active regions is, again, improved.

\begin{figure}
\begin{center}
{\resizebox{0.86\textwidth}{!}{\includegraphics[clip=]{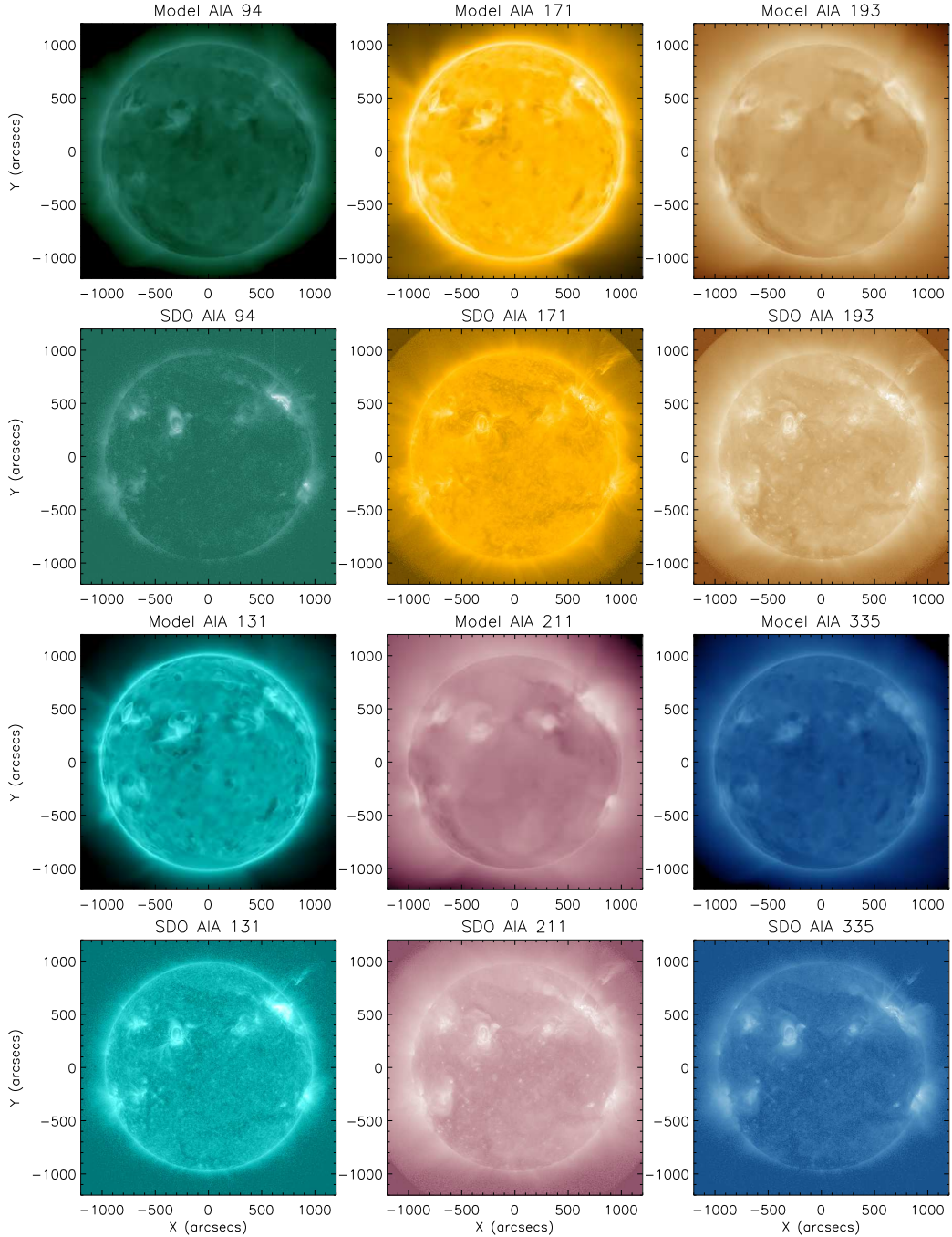}}}
\end{center}
\caption{Comparison between synthesized AIA images of the model with the
observed {\it SDO}/AIA images. Top panels (from left to right): AIA synthesized
images for $94\;{\rm \AA}$, $171\;{\rm \AA}$, and $193\;{\rm \AA}$. Panels in
second row: observational {\it SDO}/AIA images for those wavelengths. Panels in
third row: AIA synthesized images for $131\;{\rm \AA}$, $211\;{\rm \AA}$, and
$355\;{\rm \AA}$. Bottom panels: observational {\it SDO}/AIA images for
those wavelengths. The observation time is 2011 March 7 20:00$\;$UT.}
\label{fig:aia}
\end{figure}

In Figure \ref{fig:aia}, we show the comparison between the model synthesized
{\it SDO}/AIA images with the images observed by AIA \citep{lemen2012}
on board {\it SDO}. The model results are obtained with the MP5 limiter.
The wavelengths indicated at the top of each panel correspond to various
characteric temperatures. Again, the active regions are well captured.

\section{CONCLUSIONS}
\label{sec:conclusions}

We have presented our new AWSoM model. This solar
model, which is part of the SWMF, is a three-dimensional Alfv\'en wave
turbulence-driven model ranging from the upper chromosphere to the whole
heliosphere. Compared to our previous models, AWSoM includes
a generalization of the Alfv\'en wave turbulence to counter-propagating waves
on both open and closed field lines. The outward propagating waves are now
partially reflected by the Alfv\'en speed gradients and field-aligned
vorticity.  The balanced
turbulence at the apex of the closed field lines is accounted for.
We have also generalized our separate electron and ion temperature to
anisotropic ion temperatures and isotropic electron temperatures.
To distribrute the turbulence dissipation to the coronal heating of the three
temperatures, we use the results of the linear wave theory and nonlinear
stochastic heating as presented in \citet{chandran2011}. For the isotropic
electron temperature, we have now also incorporated the collisionless heat
conduction.

Our new model has no ad hoc coronal heating functions and has only a
few adjustable
parameters: three to prescribe the boundary conditions (density, temperature,
and Poynting flux of the Alfv\'en waves), a transverse correlation length
parameter for the turbulence and heat partitioning, a parameter
related to the nonlinear stochastic heating of the ions, and two parameters for
the collisionless heat conduction. Some of these parameters could potentially
be described more self-consistently. For example, the transverse correlation
length could be obtained from a time evolution equation \citep{breech2008}
instead of the simple scaling with the magnetic field strength.

Since the evolution equations of our model do not assume open or closed
field lines, those will develop self-consistently by using the data
from photospheric magnetic field observations as boundary conditions for the
magnetic field. The correctness of the
coronal heating can be tested by comparing the simulated and observed
multi-wavelength EUV images. We
performed such a validation for CR2107. We demonstrated that
our model can reproduce many features seen in the LOS images. The high
latitudinal region is somewhat distorted. This might be an artifact due to
the polynomial interpolation of the synoptic magnetogram above $60^\circ$
towards the pole. Future improvements in adapting magnetograms might address
this issue. In our companion paper, X. Meng et al. (in preparation), we
will showcase the model performance in the inner heliosphere by comparing the
results for two Carrington rotations with in-situ observations at $1\;$AU.

\acknowledgments
This work was supported by the NSF grant AGS 1322543.
W.B. Manchester IV was supported by NASA grant NNX13AG25G.
The simulations were performed on the NASA Advanced Supercomputing system,
Pleiades.

\appendix
\section{Model Simplification to Local Dissipation}
\label{sec:localdiss}

We now consider the case that the turbulence is strongly imbalanced, i.e.
$\min(w_\pm)/\max(w_\pm)\ll 1$. For simplicity, we also assume that $w_+$ is
the dominant wave, and hence, $w_-$ is the minor wave ($w_-\ll w_+$). We can
then simplify the wave equations (\ref{eq:wkbrefl}) as
\begin{equation}
  \frac{\partial w_\pm}{\partial t}
  + \nabla\cdot\left[({\bf u}\pm{\bf V}_A)w_\pm\right]
  + \frac{w_\pm}2(\nabla\cdot{\bf u})=
  \mp{\cal R}_{\rm imb}\sqrt{w_-w_+} -\Gamma_\pm w_\pm.\label{eq:imbwkbrefl}
\end{equation}
For the minor wave equation, only the right hand side of Equation
(\ref{eq:imbwkbrefl}) has terms with the dominant wave.
Hence, the reflection and dissipation term of the minor
wave equation can, in leading order of the small quantity $w_-/w_+$, be assumed
to be balanced. Using the dissipation rate $\Gamma_-= (2/L_\perp)
\sqrt{w_+/\rho}$, the minor wave energy density can then analytically be
determined as
$w_- = \frac14\rho L_\perp^2 {\cal R}_{\rm imb}^2$. Using this expression in
the dissipation rate $\Gamma_+=(2/L_\perp)\sqrt{w_-/\rho}$ of the dominant
wave equation and further noting that, to leading order, the reflection term
on the right hand side of Equation (\ref{eq:imbwkbrefl}) is much
smaller than the dissipation term for the dominant wave, we arrive at the
dominant wave equation
\begin{equation}
  \frac{\partial w_+}{\partial t}
  + \nabla\cdot\left[({\bf u}+{\bf V}_A)w_+\right]
  + \frac{w_+}2(\nabla\cdot{\bf u})= -{\cal R}_{\rm imb} w_+,
\end{equation}
valid for strongly imbalanced turbulence. We note that the resulting equation
only depends on the dominant wave energy density. A similar derivation can be
performed when the $w_-$ is the dominant wave, and we combine both cases in a
single formulation for strongly imbalanced turbulence:
\begin{equation}
  \frac{\partial w_\pm}{\partial t}
  + \nabla\cdot\left[({\bf u}\pm{\bf V}_A)w_\pm\right]
  + \frac{w_\pm}2(\nabla\cdot{\bf u})= -{\cal R}_{\rm imb} w_\pm.
  \label{eq:imb}
\end{equation}
The exponentially small minor wave energy density is now assumed to
be zero due to the
absence of a reflection term in these evolution equations (although it could
be recovered from the aforementioned analytical expression for the minor
wave). Hence, by comparing Equations (\ref{eq:imb}) and (\ref{eq:wavewkb}),
we can conclude that in strongly imbalanced turbulence, the dissipation rate is
equal to the reflection rate, i.e. the wave dissipation is local. The
dissipation does, in addition, no longer depend on the perpendicular
correlation length.
However, this derivation still dismisses the case that near the apex of closed
field lines the wave energy densities can be of equal amplitude
($w_+\approx w_-$). For the balanced turbulence, the dissipation should still
be
estimated by the original expression (\ref{eq:dissrate}). Combining
both cases results in the final expression for the wave propagation and
local dissipation
\begin{equation}\label{eq:wkblocal}
\frac{\partial w_\pm}{\partial t}+\nabla\cdot\left[({\bf u}\pm{\bf
    V}_A)w_\pm\right]+\frac{w_\pm}2(\nabla\cdot{\bf u})= -\Gamma_\pm w_\pm,
\end{equation}
in which the dissipation rate is
\begin{equation}\label{eq:disslocal}
  \Gamma_\pm = \max\left({\cal R}_{\rm imb},
  \frac{2}{L_\perp}\sqrt{\frac{w_\mp}{\rho}} \right).
\end{equation}
This approximation is valid for strongly imbalanced and balanced turbulence.
The applicability to moderately imbalanced turbulence, for example in the
transition region and chromosphere, is less certain.

\section{APPORTIONING ION AND ELECTRON HEATING}\label{sec:partition}

The partial reflection of the Alfv\'en waves due to Alfv\'en speed gradients
and field-aligned vorticity generates counter-propagating waves. The nonlinear
interaction between these oppositely directed waves results in an energy
cascade from the large scale $L_\perp$ through the inertial range to the
smaller perpendicular scales, i.e. larger perpendicular
wavenumber $k_\perp$, where it can dissipate. The apportioning of the
dissipated energy to the coronal heating functions $Q_e$, $Q_i$, and
$Q_{i\parallel}$ depends on the microphysics that is involved. In this paper,
we follow the partitioning strategy based on the dissipation of kinetic
Alfv\'en waves (KAWs) using the theory described in \citet{chandran2011}.
That formalism has the distinct advantage of providing approximated formulas
that can readily be implemented in a numerical solar wind model that is
based on the turbulent cascade of Alfv\'en waves. We have implemented those
formulas in the present model, and below, we reproduce them for convenience.

In \citet{chandran2011}, the cascading of Alfv\'en
waves transitions into cascading of KAWs. The KAWs can
dissipate when $k_\perp r_i \sim 1$, where $r_i$ is the ion gyro radius.
Among the dissipation mechanisms considered are the linear Landau damping and
linear transit time damping of KAWs, which contribute to electron and parallel
ion heating. The corresponding damping rates $\Gamma_e$ and
$\Gamma_{i\parallel}$ are
\begin{eqnarray}
  &&\Gamma_e t_c = 0.01\left( \frac{P_e}{P_i\beta_i}\right)^{1/2}\left[
    \frac{1+0.17\beta_i^{1.3}}{1+\left( 2800\beta_e\right)^{-1.25}}\right], \\
  &&\Gamma_{i\parallel} t_c = 0.08\left( \frac{P_e}{P_i}\right)^{1/4}
  \beta_i^{0.7} \exp\left(-\frac{1.3}{\beta_i}\right),
\end{eqnarray}
where $\beta_e = 2\mu_0 P_e/B^2$ and $\beta_i = 2\mu_0 P_i/B^2$ are the
electron and averaged ion plasma beta. Similar to \citet{chandran2011},
the Alfv\'en frequency $1/t_c = k_\parallel V_A$ for the parallel wavenumber
$k_\parallel$ can be rewitten as
$t_c = \rho\delta v_i^2/(\Gamma_+w_++\Gamma_-w_-)$ under the
assumption of  the critical-balance condition.
The velocity perturbation $\delta v_i$ of the Alfv\'en waves and KAWs
at $k_\perp r_i \sim 1$ is assumed to scale with $r_i/L_\perp$ via
\begin{equation}
  \rho \delta v_i^2 \approx w_d \sqrt{\frac{r_i}{L_\perp}},
\end{equation}
where $w_d=\max(w_+,w_-)$ is the dominant wave energy density. The
minor wave energy density, $w_m=\min(w_+,w_-)$, is assumed to be exponentially
small compared to $w_d$, which is, strictly speaking,
not true in the balanced turbulence regime and hence,
introduces some uncertainty in the heating partitioning.
The above scaling is compatible with the $1\;$AU observations of
\cite{podesta2007}. Furthermore, we assume similar to \citet{chandran2011},
nonlinear damping of KAWs via stochastic heating of ions,
resulting in perpendicular ion heating.
This energization is effective if $\delta v_i$ is large enough
\citep{chen2001,johnson2001}. This form of heating is the result of
stochastic ion orbits perpendicular to ${\bf B}$ in an electrostatic
potential. The damping rate for
$\beta_{i\parallel}=2\mu_0 P_{i\parallel}/B^2\lesssim 1$ is
\begin{equation}
  \Gamma_{i\perp} = 0.18 \varepsilon_i\Omega_i
  \exp\left( -\frac{h_S}{\varepsilon_i}\right),
\end{equation}
where $\Omega_i = (e/m_i)B$ is the ion gyro frequency,
$v_{i\perp} = \sqrt{2P_{i\perp}/\rho}$ is the perpendicular ion thermal speed,
$\varepsilon_i = \delta v_i/v_{i\perp}$, and $h_S$ is an input parameter for
the stochastic heating. The heating functions are 
expressed in terms of the damping rates:
\begin{eqnarray}
  && Q_e = \frac{1+\Gamma_et_c}{1+(\Gamma_e+\Gamma_{i\parallel}
    +\Gamma_{i\perp})t_c}(\Gamma_+w_++\Gamma_-w_-), \\
  && Q_{i\parallel} =\frac{\Gamma_{i\parallel}t_c}
  {1+(\Gamma_e+\Gamma_{i\parallel}+\Gamma_{i\perp})t_c}
  (\Gamma_+w_++\Gamma_-w_-), \\
  && Q_i = \Gamma_+w_++\Gamma_-w_- - Q_e.
\end{eqnarray}
The $1+$ term in these expressions is for the remaining cascading power that
succeeds to cascade to $k_\perp r_i \gg 1$, so that it can be assumed to be
dissipated via interactions with electrons and hence, contributes to electron
heating.

\section{COLLISIONLESS HEAT CONDUCTION}\label{sec:collisionless}

In this appendix, we will derive the final form of our implemented
collisionless electron heat conduction. For convenience, we will limit the
derivation to the inner heliosphere only, where the collisional heat
conduction,
Coulomb collisional heat exchange with the ions and the radiative cooling,
can be neglected in the full electron energy equation (\ref{eq:electron}).
We additionally omit the time derivate as we focus in this paper on the
steady state solar wind, and hence, Equation (\ref{eq:electron}) can be
simplified as
\begin{equation}
  \nabla\cdot\left( \frac{p_e}{\gamma - 1}{\bf u}\right)+p_e\nabla\cdot{\bf u}
  = - \nabla\cdot \left[ \frac32 \alpha p_e{\bf u}\right] + Q_e.
\end{equation}
The first term on both the left hand side and the right hand side can be
combined, resulting in
\begin{equation}
  \nabla\cdot\left( \frac{p_e}{\gamma_H - 1}{\bf u}\right)
  + p_e\nabla\cdot{\bf u}= Q_e,
\end{equation}
where we have introduced a new polytropic index $\gamma_H$ for the
electrons in the collisionless regime:
\begin{equation}
  \gamma_H = \frac{\gamma + \frac32(\gamma-1)\alpha}
        {1 + \frac32(\gamma-1)\alpha}.
\end{equation}
For our standard values $\alpha=1.05$ (taken from \citet{cranmer2009})
and $\gamma=5/3$, we obtain
$\gamma_H \approx 1.33$. By reintroducing the missing terms of Equation
(\ref{eq:electron}), we obtain our final evolution equation for the electron
pressure:
\begin{equation}
  \frac{\partial}{\partial t}\left(\frac{P_e}{\gamma_e-1}\right)
  +\nabla\cdot\left(\frac{P_e}{\gamma_e-1}{\bf
    u}\right)+P_e\nabla\cdot{\bf u}=
  -\nabla\cdot{\bf q}_e^*
  +\frac{N_ik_B}{\tau_{ei}}\left(T_i-T_e\right)-Q_{\rm rad}+ Q_e.
\label{eq:electronpressuregamma}
\end{equation}
where
\begin{equation}
  \gamma_e = \gamma f_S + \gamma_H (1-f_S)
\end{equation}
interpolates the electron polytropic index $\gamma_e$ between the
collisional regime where $\gamma_e=\gamma$
and the collisionless regime where $\gamma_e=\gamma_H$ and the interpolation
function $f_S$ is taken to be the same as Equation (\ref{eq:interpolation}).
The $*$ in ${\bf q}_e^*$ indicates that we set ${\bf q}_{e,H}=0$ in the
electron heat flux (\ref{eq:heatflux}), i.e. ${\bf q}_e^* = f_S {\bf q}_{e,S}$,
since it is now parameterized via a spatially varying $\gamma_e$.

The main difference between Equation
(\ref{eq:electronpressuregamma}) and (\ref{eq:electron}) is the use of
$\gamma_e$ instead of $\gamma$ in the time derivative, and hence, the time
evolution of both (ad hoc) formulations is different.
As a final note, we mention that a spatially varying electron polytropic index
does not negatively impact the shock evolution of coronal mass ejections
(CMEs). Since the electron speed of sound is much larger than the CME speeds,
only the ions will be heated by the CME shock. The ion fluid still uses the
standard polytropic index $\gamma=5/3$.

\newpage

\begin{deluxetable}{l|l}
\tablecaption{The model parameters.
\label{table:parameters}}
\startdata
Parameter & Value \\
\hline
$N_\odot$ & $2\times 10^{17}\;{\rm m}^{-3}$ \\
$T_\odot$ & $50,000\;{\rm K}$ \\
$(S_A/B)_\odot$ & $1.1\times 10^6\;{\rm W}\,{\rm m}^{-2}\,{\rm T}^{-1}$ \\
$L_\perp\sqrt{B}$ & $1.5\times 10^5\;{\rm m}\,\sqrt{\rm T}$ \\
$h_S$ & 0.17 \\
$\alpha$ & 1.05 \\
$r_H$ & $5\;R_\odot$
\enddata
\end{deluxetable}

\newpage

\newpage

\newpage

\newpage

\newpage

\end{document}